\documentclass[10pt,conference]{IEEEtran}
\usepackage{cite}
\usepackage{amsmath,amssymb,amsfonts}
\usepackage{algorithm}
\usepackage{graphicx}
\usepackage{svg}
\usepackage{subcaption}
\usepackage{textcomp}
\usepackage{xcolor}
\usepackage[hyphens]{url}
\usepackage{fancyhdr}
\usepackage{hyperref}
\usepackage[noend]{algpseudocode}
\usepackage{pifont}
\algnewcommand\algorithmicinput{\textbf{Input:}}
\algnewcommand\INPUT{\item[\algorithmicinput]}
\algnewcommand\algorithmicoutput{\textbf{Output:}}
\algnewcommand\OUTPUT{\item[\algorithmicoutput]}
\algnewcommand\algorithmicparameters{\textbf{Hyperparameter:}}
\algnewcommand\PARAMETER{\item[\algorithmicparameters]}
\newtheorem{definition}{Definition}
\newcommand{\cmark}{\textcolor{green}{\checkmark}}
\newcommand{\xmark}{\textcolor{red}{\ding{55}}}

\newcommand{\hpcayear}{2027}

\title{\sys: Padding is Simple and Efficient \\ for Deterministic LLM Inference}

\newcommand\hpcaauthors{Shiju Zhao$^*$, Jiacheng Yang$^\dagger$, Qihang Chen$^\dagger$, Junhao Hu$^{*\ddagger}$, Jiaqi Zheng, Guihai Chen, Xusheng Chen$^\dagger$}
\newcommand\hpcaaffiliation{State Key Laboratory for Novel Software Technology, Nanjing University, China \\ $^\dagger$Hunyuan Team, Tencent, China \\ $^\ddagger$School of Computer Science, Peking University, China}
\newcommand\hpcaemail{Email(s):}
\newcommand{\sys}{\textsc{CoRun}}

\author{
  \ifdefined\hpcacameraready
    \IEEEauthorblockN{\hpcaauthors{}}
      \IEEEauthorblockA{
        \hpcaaffiliation{} \\
        \hpcaemail{}
      }
  \else
    \IEEEauthorblockN{\hpcaauthors{}}
      \IEEEauthorblockA{
        \hpcaaffiliation{}
      }
  \fi 
}

\fancypagestyle{camerareadyfirstpage}{%
  \fancyhead{}
  
  \fancyhead[C]{
    \ifdefined\aeopen
    \parbox[][12mm][t]{13.5cm}{\hpcayear{} IEEE International Symposium on High-Performance Computer Architecture (HPCA)}    
    \else
      \ifdefined\aereviewed
      \parbox[][12mm][t]{13.5cm}{\hpcayear{} IEEE International Symposium on High-Performance Computer Architecture (HPCA)}
      \else
      \ifdefined\aereproduced
      \parbox[][12mm][t]{13.5cm}{\hpcayear{} IEEE International Symposium on High-Performance Computer Architecture (HPCA)}
      \else
      \parbox[][0mm][t]{13.5cm}{\hpcayear{} IEEE International Symposium on High-Performance Computer Architecture (HPCA)}
    \fi 
    \fi 
    \fi 
    \ifdefined\aeopen 
      \includegraphics[width=12mm,height=12mm]{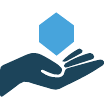}
    \fi 
    \ifdefined\aereviewed
      \includegraphics[width=12mm,height=12mm]{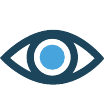}
    \fi 
    \ifdefined\aereproduced
      \includegraphics[width=12mm,height=12mm]{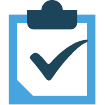}
    \fi
  }
  \fancyfoot[C]{}
}
\begin{document}
\maketitle

\ifdefined\hpcacameraready 
  \thispagestyle{camerareadyfirstpage}
  \pagestyle{empty}
\else
  \thispagestyle{plain}
  \pagestyle{plain}
\fi

\newcommand{\hpcaheight}{0mm}
\ifdefined\eaopen
\renewcommand{\hpcaheight}{12mm}
\fi

\renewcommand{\thefootnote}{\fnsymbol{footnote}}
\footnotetext[1]{This work was completed during his internship at Tencent.}
\renewcommand{\thefootnote}{\arabic{footnote}}


\begin{abstract}

  Despite fixed sampling parameters and random seeds, Large Language Model (LLM) inference exhibits output inconsistency, which undermines downstream tasks such as model evaluation and reinforcement learning.  A major source of this nondeterminism is batch-dependent GPU execution: dynamic input shapes change kernel tiling and floating-point reduction orders. Existing systems address this problem with batch-invariant kernels, but these kernels restrict optimized tiling and split reductions, increasing more than 2$\times$ latency and reducing serving throughput by up to 74\%. This paper observes that although most kernels are not batch-invariant, they are \emph{position-invariant}. Leveraging this property, we present \sys, a scheduling-based system that achieves deterministic inference without requiring batch invariance. \sys\ employs isolated prefill and fixed-shape batched decode to handle the two stages of LLM inference, respectively, leveraging CUDA graphs for efficient execution and simplified implementation. Experiments on LLMs with diverse architectures, including Qwen and DeepSeek, show that \sys\ ensures determinism while improving throughput by 15--324\% over batch-invariant approaches, reducing time-to-first-token by 51.8\% and time-per-output-token by 48.6\% on average.

\end{abstract}

\section{Introduction}

Large language model (LLM) inference can produce different token sequences for
the same prompt even when users fix the sampling parameters and random
seed~\cite{anadkat2023seed, atil2025non, chann2023non, yuan2025fp32}. The discrepancy is easy
to overlook because it often begins as a small numerical perturbation, but
autoregressive generation can amplify it into a different answer. In hosted
models, repeated nominally deterministic runs have produced task-score gaps of
up to 70\% between the best and worst observed outcomes~\cite{atil2025non}.
Such variation makes model comparisons sensitive to incidental serving
conditions. It also complicates debugging and reproducibility, and can create
rollout mismatches that harm Reinforcement Learning (RL)
training~\cite{feng2025offpolicy, he2025nondeterminism, zhong2026diagnosing}. Deterministic execution
is consequently becoming an explicit requirement in model evaluation, debugging, stability analysis, and consistent post-training~\cite{deepseekai2026deepseekv4,he2025nondeterminism}.

A major source of this nondeterminism is dynamic GPU batching. Modern LLM
servers admit and retire requests between model iterations to increase
utilization~\cite{yu2022orca,kwon2023efficient}. The resulting batch shape
depends on request arrival times, prompt lengths, and generation progress.
GPU libraries use these dimensions to choose tiling and parallel reduction
strategies. Changing the shape can change the order of floating-point
operations in Mean, RMSNorm, Softmax, matrix multiplication (MatMul), and attention~\cite{goldberg1991floating,dao2022flashattention,
dao2024flashattention2}. Since floating-point addition is not associative, the
same token can receive different low-order bits when it runs with different
co-runners. Section~\ref{sec:motivation} explains this mechanism in detail using an
illustrative reduction kernel.

Existing deterministic inference stacks address this problem with
\emph{batch-invariant} kernels~\cite{deepseekai2026deepseekv4, he2025nondeterminism, sglang2025det,
vllm2026det}. A batch-invariant kernel returns the same result for a token
regardless of both the batch size and the token's position. It achieves this
property by retaining a fixed tiling and reduction order across supported
shapes. The restriction prevents the kernel from selecting the most efficient
decomposition for each input and can disable optimizations such as Split-K and
Split-KV.
\begin{figure}
    \centering
    \begin{subfigure}[t]{0.49\linewidth}
        \includegraphics[width=\linewidth]{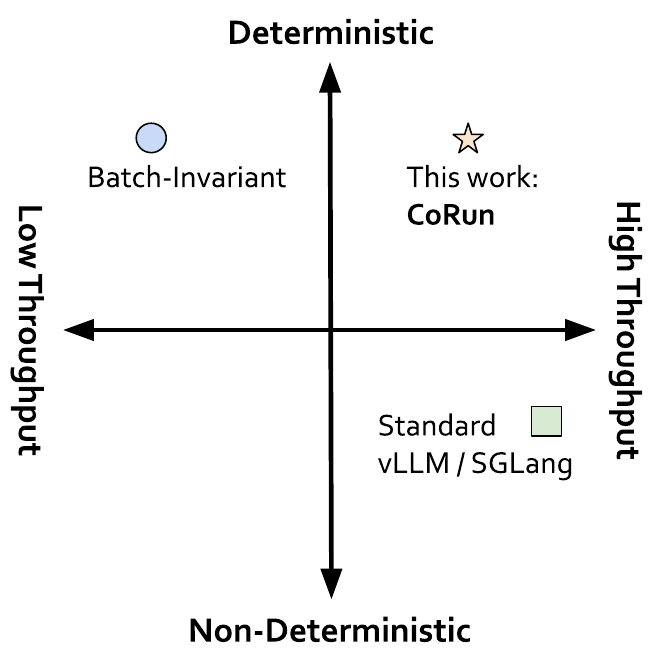}
        \caption{Design space}
        \label{fig:4a}
    \end{subfigure}
    \begin{subfigure}[t]{0.49\linewidth}
        \includegraphics[width=\linewidth]{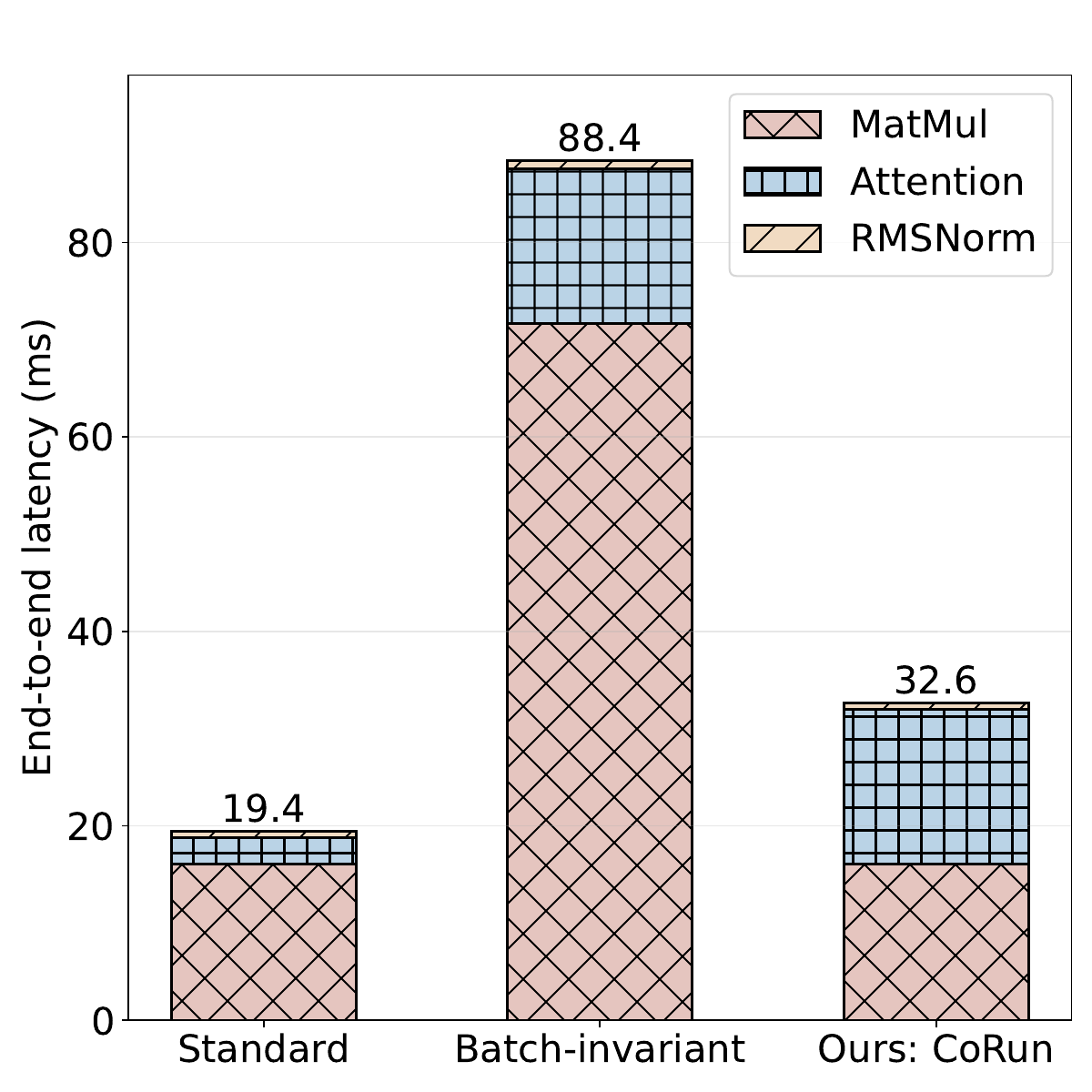}
        \caption{Latency breakdown}
        \label{fig:4b}
    \end{subfigure}
    \caption{\sys's design objective and the latency opportunity. (a) Positioning in the determinism-throughput design space. (b) Decode-step latency breakdown for a batch of 32 requests with variable context lengths.}
    \label{fig:4}
\end{figure}

\figurename~\ref{fig:4a} summarizes the resulting design space. Standard vLLM and
SGLang provide high throughput but can produce batch-dependent outputs, while
batch-invariant execution restores determinism at lower throughput. This work
targets the upper-right region by retaining fast standard kernels and
controlling the shapes on which they execute. \figurename~\ref{fig:4b} illustrates
the opportunity with the latency breakdown of one variable-length decode
step. Batch-invariant execution increases end-to-end latency from 19.4\,ms to
88.4\,ms, with MatMul accounting for most of the increase. \sys\ reduces the
latency to 32.6\,ms by avoiding batch-invariant implementations for most
operators, including MatMul and RMSNorm.

\begin{figure}
    \centering
    \begin{subfigure}[t]{0.49\linewidth}
        \includegraphics[width=\linewidth]{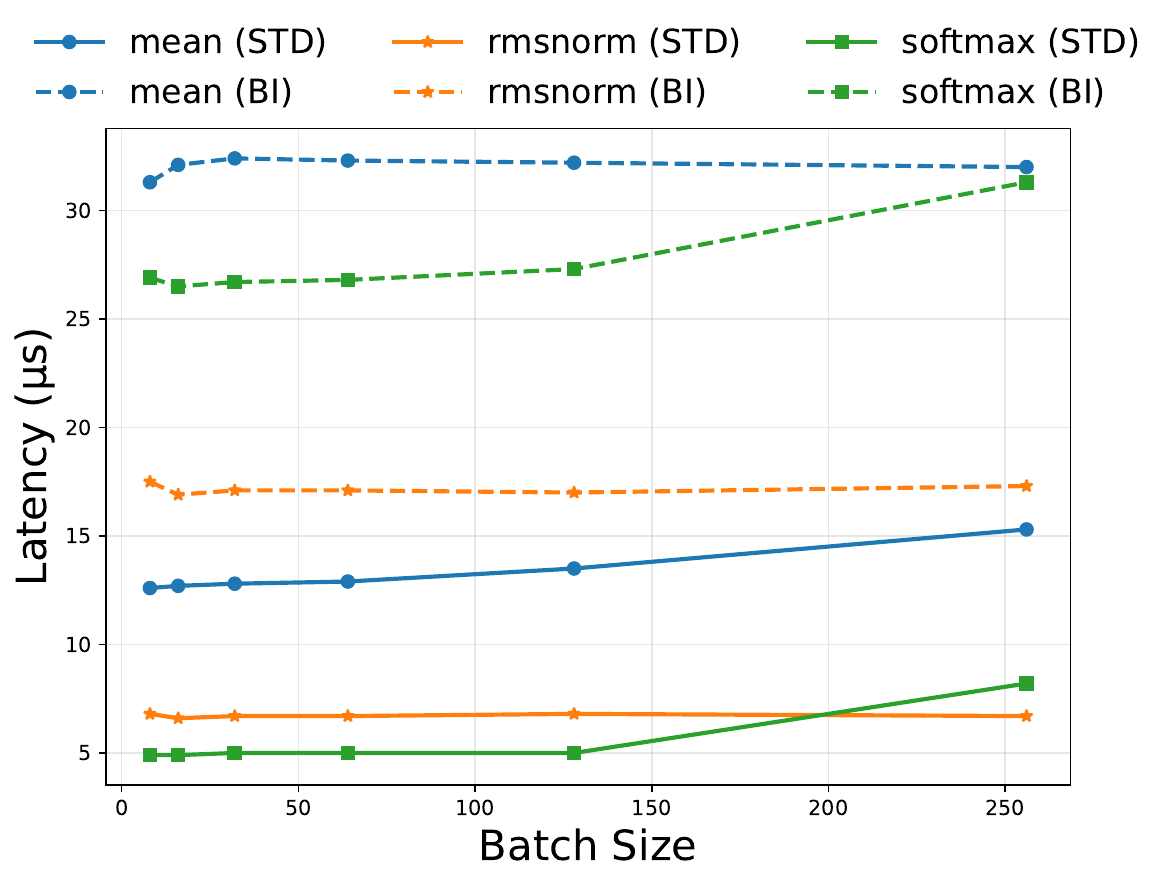}
        \caption{Mean, RMSNorm, SoftMax}
        \label{fig:2a}
    \end{subfigure}
    \begin{subfigure}[t]{0.49\linewidth}
        \includegraphics[width=\linewidth]{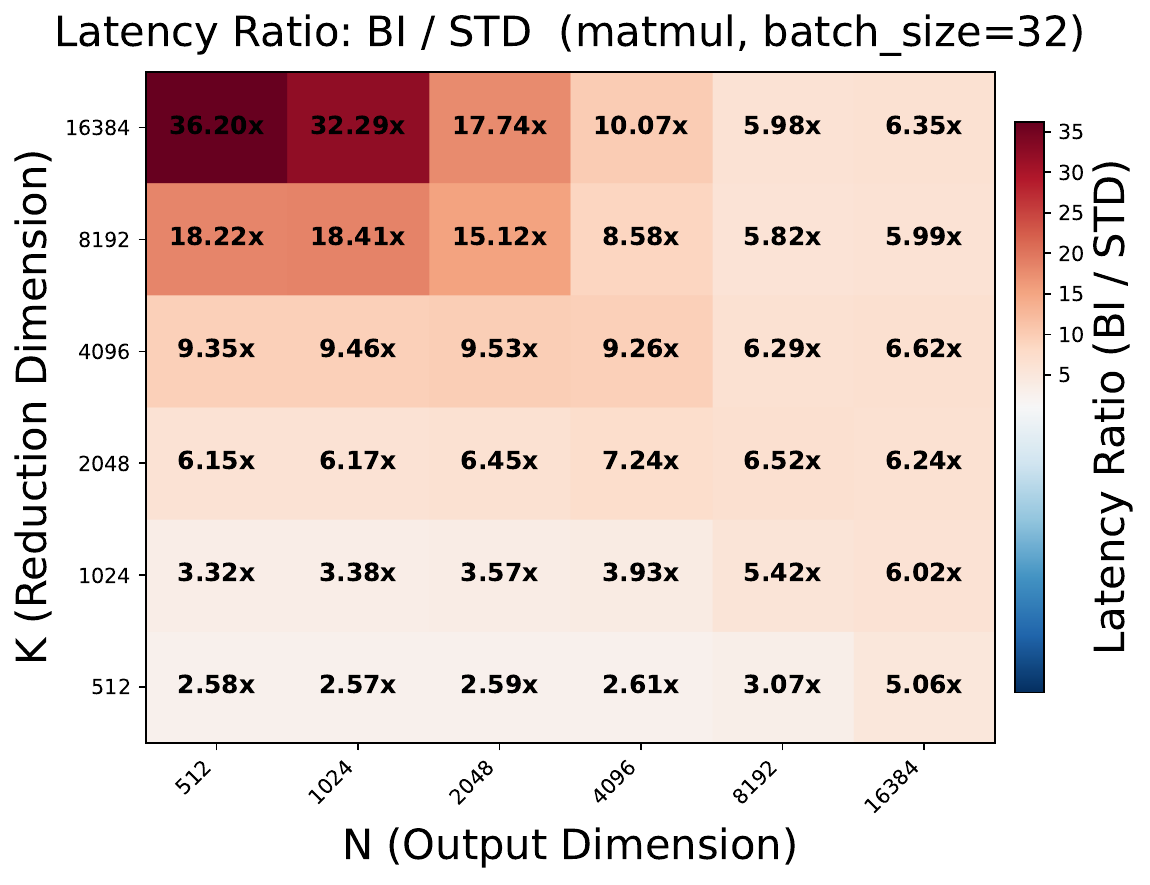}
        \caption{MatMul}
        \label{fig:2b}
    \end{subfigure}
    \caption{Performance comparison between standard (STD) non-deterministic kernels and batch-invariant (BI) kernels.}
    \label{fig:2}
\end{figure}

The operator-level cost is also substantial. \figurename~\ref{fig:2} compares standard
(STD) and batch-invariant (BI) implementations of four common operators. Each
point is the mean of 100 executions. Across batch sizes, BI increases the
latency of Mean, RMSNorm, and Softmax by 2.12--5.61$\times$
(\figurename~\ref{fig:2a}). For matrix multiplication, the overhead grows with the
reduction dimension and reaches 36.2$\times$ (\figurename~\ref{fig:2b}). A larger
output dimension exposes more independent work and narrows the gap, but does
not remove it. End to end, these slower operators reduce serving throughput by
up to 74\% (Section~\ref{sec:throughput}).

This paper observes that full batch invariance is stronger than necessary.
Although the optimized kernels in our model stack are generally not
batch-invariant, most are \emph{position-invariant}: a token retains the same
result when it moves within a batch of the same shape. This weaker property
suggests a different route to determinism: control the execution shape seen by
each request, then retain the faster position-invariant kernels. We formalize
the distinction and characterize the relevant kernels in
Section~\ref{sec:motivation}.

Applying this idea to existing LLM serving systems is challenging because prefill and decode\footnote{Prefill and decode are two stages of LLM inference (Section~\ref{sec:background}).}
have different shape distributions. Prefill processes a variable-length
prompt, so padding every request to one global shape would waste substantial
computation. Batching several prompts, however, makes the shape of each
prefill depend on its co-runners. Decode contributes one query token per active
request, but the active set changes whenever a request arrives or completes.
Sampling creates a final complication: a request must consume the same random
number after it moves to another batch slot.

We present \sys, a scheduling-based system for deterministic, high-throughput
LLM inference. \sys\ isolates each request during prefill, preserving its
natural prompt shape while removing co-runner dependence. During decode, it
pads the active batch to the maximum concurrency and replays one fixed-size
CUDA Graph. \sys\ disables attention Split-KV for decode and uses deterministic
collectives for the small set of operators that are not position-invariant.
The sampler binds Random Number Generator (RNG) state to requests rather than batch slots and pads its
input to the same fixed shape. Together, these mechanisms make the output
independent of co-running requests, batch position, and arrival order without
requiring every operator kernel to be batch-invariant. We formalize this guarantee
as an end-to-end determinism contract and prove it by induction over
autoregressive generation. Isolated prefill establishes an identical initial
request state; fixed-shape decode and position-invariant kernels preserve that
state across batch compositions and positions; request-bound RNG state then
ensures that identical logits produce identical sampled tokens.

We make the following contributions:
\begin{itemize}
  \item We distinguish batch invariance from position invariance, characterize
  the invariance properties and costs of common LLM kernels, and identify
  execution-shape control as an alternative to making the full kernel stack
  batch-invariant.
  \item We design \sys, which combines isolated prefill, fixed-shape
  CUDA-Graph decode, selective kernel constraints, and request-bound sampling
  state in an existing continuous-batching engine. We state its determinism
  contract and prove that these mechanisms produce the same token sequence
  across co-running requests, batch positions, and arrival orders.
  \item We evaluate \sys\ on Qwen3-235B-A22B, DeepSeek-V3, and Hy3 using
  production, code-editing, and synthetic workloads. \sys\ produces identical
  outputs in our determinism suite and improves throughput by 15--324\% over
  batch-invariant inference, more than doubling aggregate throughput for every
  evaluated model.
\end{itemize}

\section{Background}\label{sec:background-section}
\subsection{LLM Inference and Continuous Batching}\label{sec:background}
An autoregressive LLM serves a request in two stages. The \emph{prefill}
stage processes all prompt tokens in parallel, materializes their key-value
(KV) cache, and produces the first output token. The \emph{decode} stage then
generates one token per request in each model iteration, reusing the cached
keys and values. Prefill therefore exposes substantial parallelism but may
contain thousands of tokens, whereas each decode iteration contributes only
one new token per active request.

Modern serving systems batch requests at iteration granularity rather than waiting
for an entire request to finish~\cite{yu2022orca,kwon2023efficient}. This
\emph{continuous batching} policy admits newly arrived requests and removes
completed requests between iterations, improving GPU utilization and avoiding
head-of-line blocking. The resulting batch is dynamic: its request count,
token count, sequence-length distribution, and token-to-slot mapping can all
change from one iteration to the next. Systems can also mix prefill tokens and
decode tokens in one iteration. Sarathi-Serve, for example, divides a long
prefill into bounded chunks (named chunked prefill) and combines them with ongoing decodes to reduce
generation stalls~\cite{agrawal2024sarathi}. These scheduling techniques
improve serving efficiency, but they make the GPU execution shape depend on
the requests that happen to run together.

\subsection{CUDA Graphs}

A conventional GPU execution path launches each kernel separately, incurring
CPU, runtime, and driver overhead for every launch. A CUDA Graph records a
sequence of kernels, memory operations, and their dependencies, instantiates
that sequence once, and replays the entire graph with one launch
operation~\cite{nvidia2026cudagraphs}. Replay amortizes setup and dispatch
overhead, which is particularly valuable for decode iterations composed of
many short kernels.

Graph replay expects the captured operations, tensor addresses, and launch
configuration to remain valid. LLM engines therefore capture graphs for a set
of sequence lengths and pad a smaller runtime batch to the smallest captured size
that can contain it~\cite{vllm2026cudagraph}. For example, an iteration with 31
query tokens can use a graph captured for 32 tokens by adding one masked
padding token. The redundant computation is often cheaper than returning to
the per-kernel eager path.

Decode is well suited to graph capture because every active request contributes
one query token and the batch size is bounded by the maximum concurrency.
Prefill shapes vary over a much wider range and include long prompts or prompt
chunks. Capturing enough prefill graphs would consume substantial memory,
while padding a short prompt to a large capture size would waste significant
computation. Serving engines therefore commonly use CUDA Graphs for decode
and an eager path for prefill. 
This asymmetry suggests separating variable-shaped prefill from fixed-shape decode, an opportunity we exploit in the rest of this paper.
\section{Motivation}\label{sec:motivation}

Continuous batching makes the execution shape depend on the requests that
happen to run together, as discussed above.
This section first explains why shape changes make standard GPU reductions
batch-dependent. It then contrasts batch invariance with position invariance
and shows why the latter creates an opportunity for a faster deterministic
inference system.

\subsection{Batch-Dependent GPU Reductions}

\begin{figure}
    \centering
    \begin{subfigure}[t]{\linewidth}
        \includegraphics[width=\linewidth]{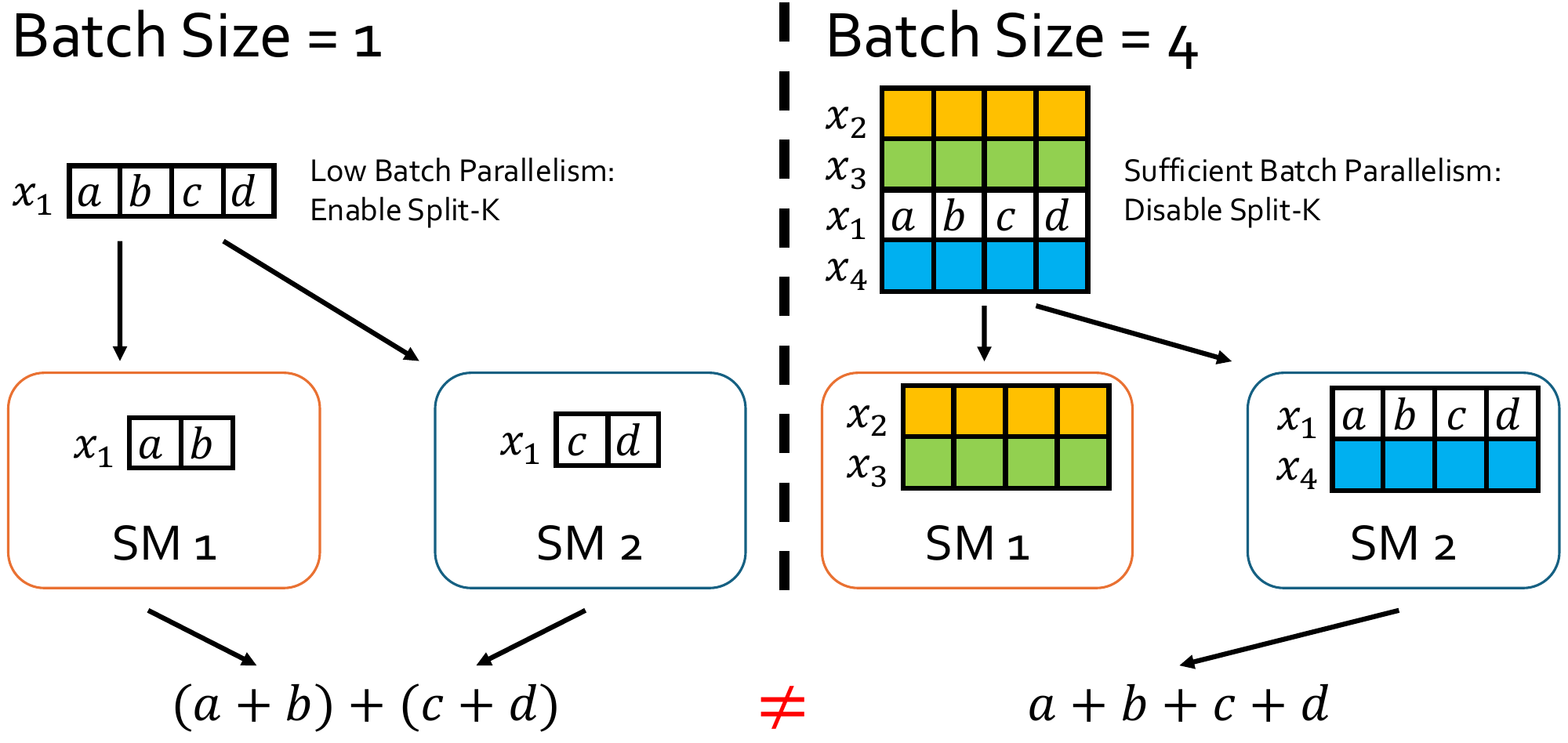}
        \caption{Adaptive tiling in a batch-dependent kernel}
        \label{fig:3a}
    \end{subfigure}
    \vskip 0.1in
    \begin{subfigure}[t]{\linewidth}
        \includegraphics[width=\linewidth]{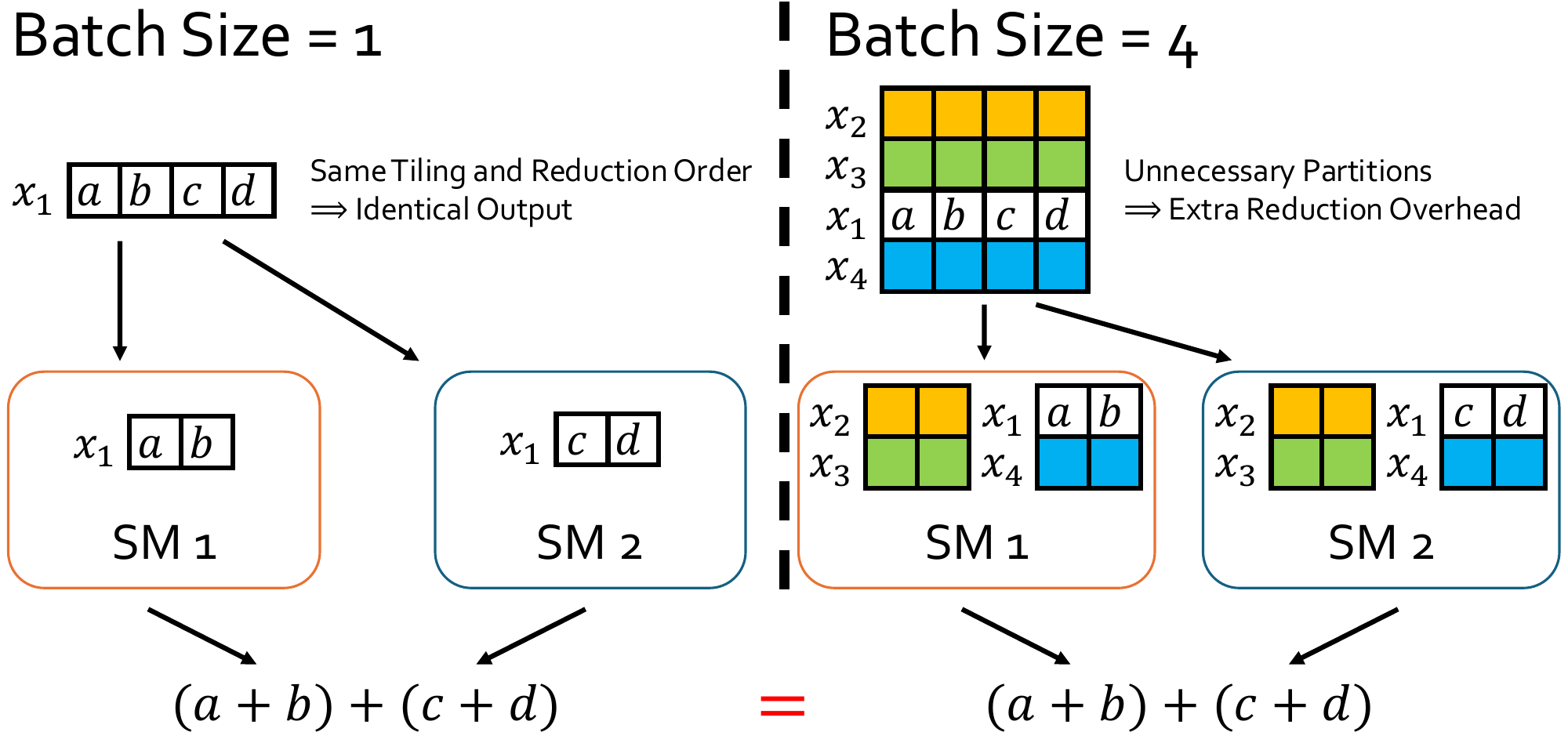}
        \caption{Fixed tiling in a batch-invariant kernel}
        \label{fig:3b}
    \end{subfigure}
    \caption{Adaptive and fixed tiling for a sum reduction. Adaptive tiling changes the reduction tree with batch size, whereas fixed tiling preserves the tree at the cost of extra work.}
    \label{fig:3}
\end{figure}

GPU kernels divide their inputs into tiles that thread blocks process in
parallel. Kernel libraries adapt this tiling to the input shape to improve
locality and expose enough parallel work to occupy the GPU. Consequently, the
tiling selected for one token can depend on the other tokens in its batch.
\figurename~\ref{fig:3} illustrates this behavior with a sum kernel. Each row
$x_i$ contains four values to reduce, and $x_1=[a,b,c,d]$ is the target input
whose output we compare across batch sizes.

In \figurename~\ref{fig:3a}, a batch of one row provides little parallelism.
The adaptive kernel therefore splits the reduction dimension across two
Streaming Multiprocessors (SMs). The two SMs compute $a+b$ and $c+d$
independently, and then combine the partial sums as $(a+b)+(c+d)$. With four
rows, row-level parallelism is sufficient to occupy both SMs. The kernel no
longer splits each row and instead assigns complete rows to the SMs; the
target row is reduced locally in the order represented by $a+b+c+d$. This
adaptive choice improves utilization, but makes the kernel
\emph{batch-dependent}: changing the batch size changes the partitioning and
reduction tree used for the same row.

The two trees are equivalent over real numbers but need not be equivalent in
floating-point arithmetic. Addition is not associative:
\[
  \operatorname{fl}(\operatorname{fl}(a+b)+c)
  \ne
  \operatorname{fl}(a+\operatorname{fl}(b+c)),
\]
because each operation rounds its result to a finite
representation~\cite{goldberg1991floating}. Consequently, two legal reduction
trees can produce different low-order bits for $x_1$.

A batch-invariant kernel removes this dependence by using a canonical
partition and reduction tree for every supported batch size.
\figurename~\ref{fig:3b} always divides each row into $[a,b]$ and $[c,d]$ and
combines the two partial sums in the same order. The target row therefore
produces $(a+b)+(c+d)$ at both batch sizes. This fixed strategy sacrifices
adaptivity: at batch size four, row-level parallelism is already sufficient,
yet the kernel still creates partial sums and performs an additional
reduction. A kernel could instead prohibit splitting at every batch size, but
that choice would leave the GPU underutilized for a batch of one. Batch
invariance must choose one shape-independent strategy and thus cannot optimize
both cases.

The sum kernel represents a broader class of GPU reductions. Matrix
multiplication may partition its reduction dimension $K$ across thread blocks,
a strategy called \emph{Split-K}. Attention kernels similarly use
\emph{Split-KV} to partition a request's KV sequence when the batch provides
insufficient query parallelism~\cite{dao2022flashattention,
dao2024flashattention2,dao2023flashdecoding}. In each case, changing the batch
shape can change the tile size, partition count, thread-block assignment, and
floating-point reduction order. The resulting numerical difference is
usually small, but autoregressive generation repeatedly feeds each sampled
token back into the model. A small logit perturbation can eventually change a
token decision and cause two sequences to diverge~\cite{yuan2025fp32,
he2025nondeterminism}.

\begin{figure}
    \centering
    \includegraphics[width=\linewidth]{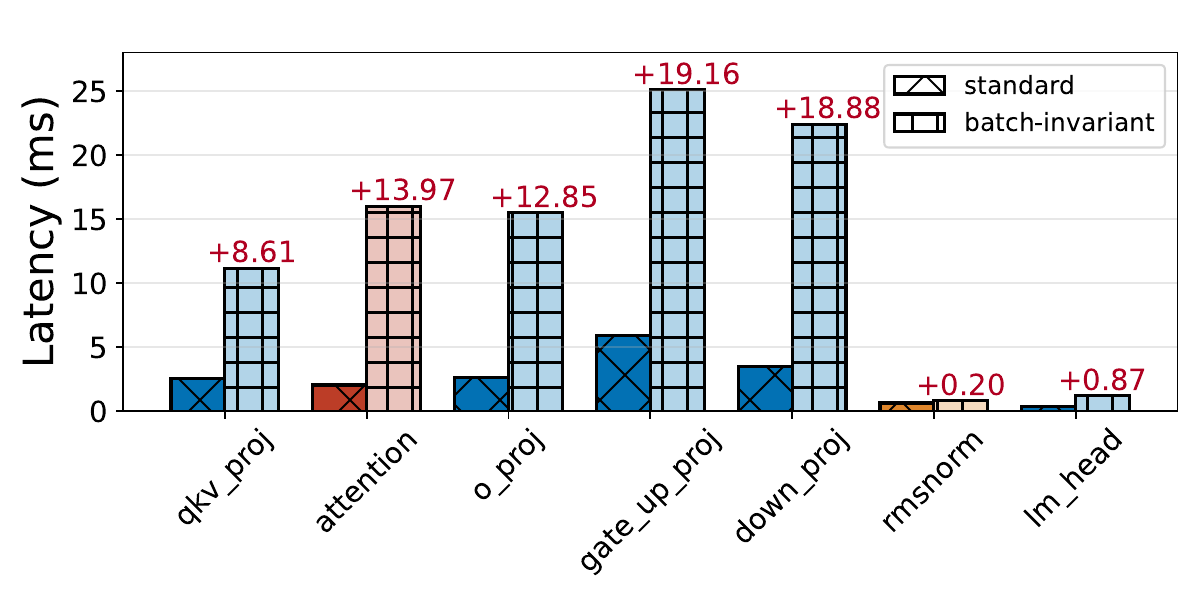}
    \caption{Per-forward total latency of different operators.}
    \label{fig:op}
\end{figure}

\figurename~\ref{fig:op} quantifies this cost for one decode forward pass of
Qwen3-235B-A22B. We aggregate the latency of each operator type across all
model layers and compare the standard and batch-invariant implementations.
Batch invariance increases the latency of every measured operator, but the
projection and attention kernels dominate the overhead. The four MatMul
groups add 8.61--19.16\,ms each, and batch-invariant attention adds
13.97\,ms. In contrast, RMSNorm and the language-model head (lm\_head) add only
0.20\,ms and 0.87\,ms, respectively. These results show that the performance
loss is distributed across the forward pass but is concentrated in its
compute-intensive MatMul kernels. Replacing those kernels with
their standard position-invariant implementations therefore offers the
largest opportunity to recover deterministic inference performance.

Most standard token-wise kernels satisfy the weaker property of position
invariance. For a fixed shape, the kernel selects the same tile dimensions,
thread-block layout, and reduction partitions for every batch position.
Moving $x_1$ to another row changes its address or program identifier, but
does not change how its four values are partitioned or combined. The output
therefore remains unchanged. Prior work calls this behavior a
\emph{shape-consistent reduction} and observes it in common LLM
operators~\cite{he2025nondeterminism,gond2026llm42}.

\subsection{Batch and Position Invariance}

\begin{figure}
    \centering
    \begin{subfigure}[t]{\linewidth}
        \includegraphics[width=\linewidth]{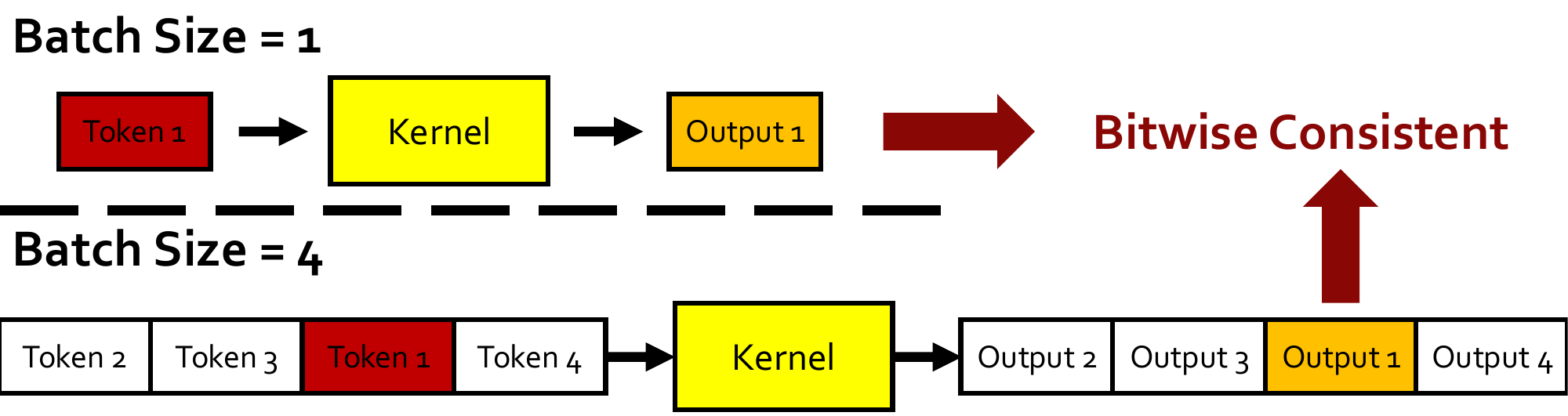}
        \caption{Batch-invariant kernel}
        \label{fig:1a}
    \end{subfigure}
    \begin{subfigure}[t]{\linewidth}
        \includegraphics[width=\linewidth]{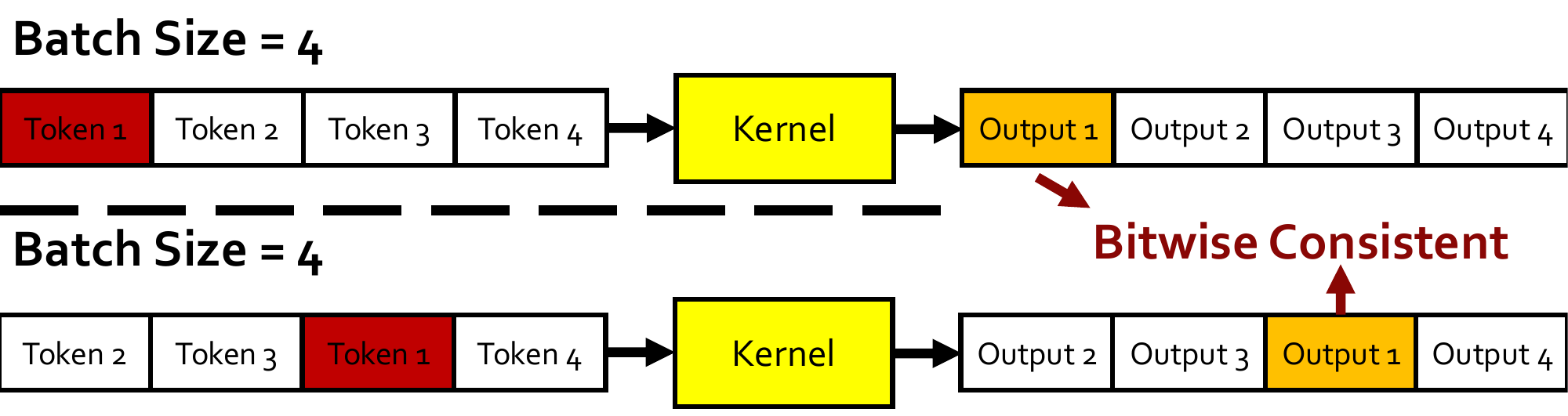}
        \caption{Position-invariant kernel}
        \label{fig:1b}
    \end{subfigure}
    \caption{Illustration of batch invariance and position invariance.}
    \label{fig:1}
\end{figure}
The tiling example explains the numerical mechanism, while
\figurename~\ref{fig:1} states the two invariance properties from a serving
perspective. \figurename~\ref{fig:1a} compares the same target token across
different batch shapes and positions. A batch-invariant kernel must preserve
its output in every case. \figurename~\ref{fig:1b} holds the batch shape fixed and
only moves the target token. A position-invariant kernel needs to preserve the
output under this weaker change.

Consider a kernel $F_n$ that processes an ordered batch
$X=(x_1,\ldots,x_n)$. For kernels with request-local metadata, such as KV
lengths and page tables, let $x_i$ include that metadata. 
\begin{definition}
The kernel $F$ is \emph{batch-invariant} if a request receives the same output under arbitrary
batch compositions and positions:
\[
  x_i=x'_j
  \ \Longrightarrow\
  F_n(X)_i = F_m(X')_j ,
  \qquad \forall n,m,i,j.
\]
\end{definition}

This property is stronger than deterministic execution at one shape: the
request output must remain unchanged when either the batch size or its
co-runners change.
\begin{definition}
The kernel $F$ is \emph{position-invariant} if it is equivariant to permutations
at a fixed shape. For every permutation matrix $P$,
\[
  F_n(PX)=P F_n(X).
\]
\end{definition}

Position invariance allows a request to move between batch slots, but requires
the overall execution shape to remain fixed. Batch invariance implies position
invariance, but the converse does not hold. In particular, a reduction kernel
may choose a different tile decomposition for $F_n$ and $F_m$ while applying
the same decomposition to every position within $F_n$.

Existing deterministic inference stacks obtain batch invariance by fixing
tile sizes and reduction trees across supported shapes~\cite{deepseekai2026deepseekv4, he2025nondeterminism,
sglang2025det,vllm2026det}. A fixed decomposition prevents co-runners from
changing the arithmetic order, but it also prevents the kernel from selecting
the fastest decomposition for each shape. It may reduce parallelism, add
passes over intermediate results, or disable Split-K and Split-KV. Tree-Based
Invariant Kernels (TBIK) apply the same principle across tensor-parallel
configurations by aligning intra- and inter-GPU reductions through a fixed
hierarchical tree~\cite{zhang2026deterministic}.

\begin{scriptsize}
\begin{table}[!ht]
  \centering
  \caption{Invariance properties of the kernel implementations considered in
  this work. Fixed-tree AllReduce uses a fixed reduction topology.}
  \label{tab:kernel-invariance}
  \begin{tabular}{lcc}
    \hline
    \textbf{Kernel} & \textbf{Batch-Invariant} & \textbf{Position-Invariant} \\
    \hline
    \hline
    Mean & \xmark & \cmark \\
    RMSNorm & \xmark & \cmark \\
    SoftMax & \xmark & \cmark \\
    MatMul & \xmark & \cmark \\
    Ring AllReduce & \xmark & \xmark \\
    Fixed-Tree AllReduce & \cmark & \cmark \\
    FlashAttention w/ Split-KV & \xmark & \xmark \\
    FlashAttention w/o Split-KV & \cmark & \cmark \\
    \hline
  \end{tabular}
\end{table}
\end{scriptsize}

\tablename~\ref{tab:kernel-invariance} summarizes the implementations used in
our stack~\cite{gond2026llm42,zhang2026deterministic}. Ring AllReduce and
Split-KV attention remain position-dependent because tensor offsets or KV
partitions can follow different reduction orders. \sys\ handles these
exceptions with deterministic collectives and by disabling decode-time
Split-KV, while retaining faster position-invariant implementations for the
remaining kernels.

The key opportunity is therefore to make each inference stage present a
repeatable shape to the model. Position-invariant kernels can then retain
their optimized, shape-specific tiling without allowing batch position or
co-runners to change a request's output. The next section turns this
observation into \sys's isolated-prefill, fixed-shape-decode, and
shape-aligned-sampling design.

\section{Design}

\begin{figure}
    \centering
    \includegraphics[width=\linewidth]{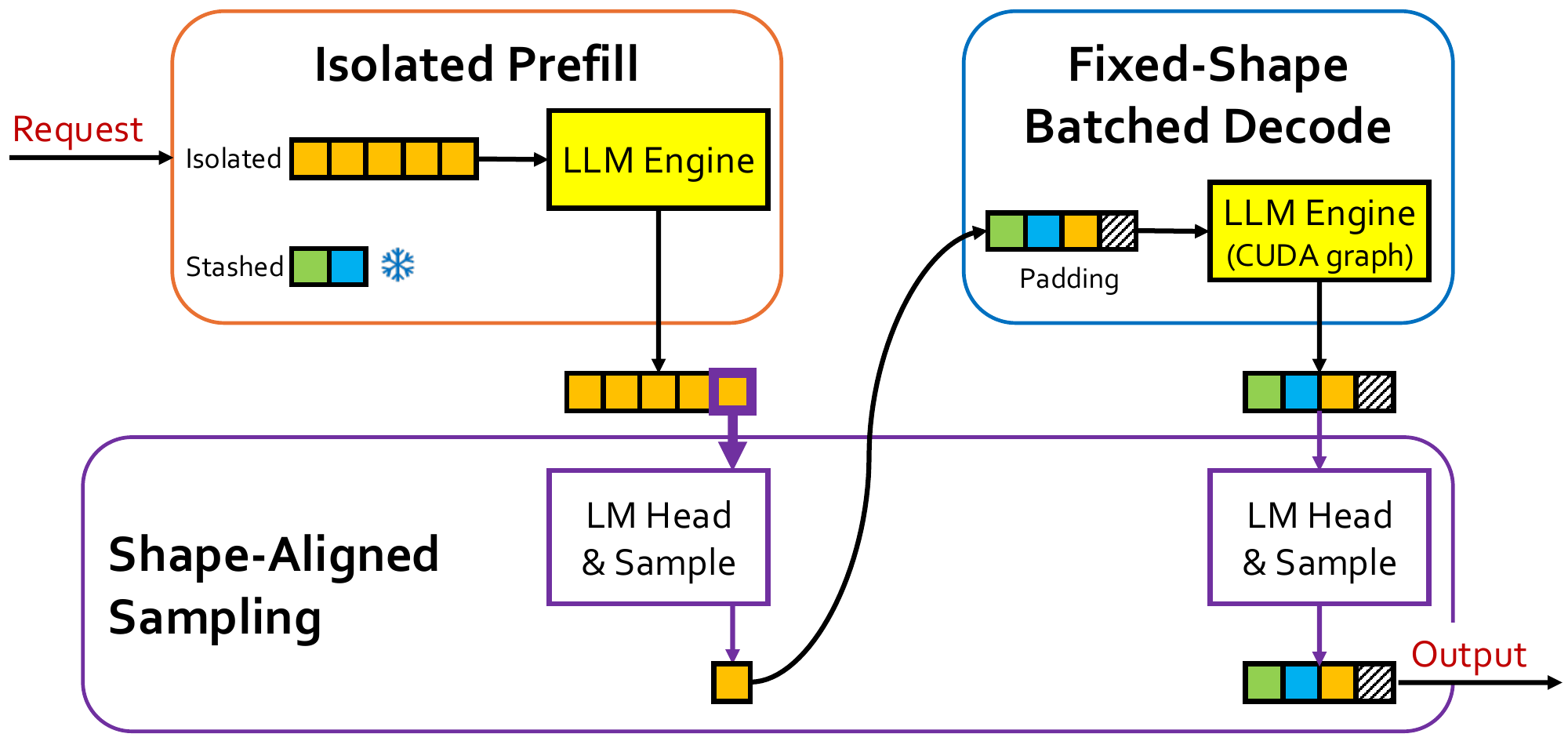}
    \caption{\sys\ overview. Each square represents a token; colors distinguish
    requests.}
    \label{fig:sys}
\end{figure}

\sys\ builds on the opportunity identified in
Section~\ref{sec:motivation}: it controls the execution context seen by each
request instead of replacing every optimized operator with a batch-invariant implementation.
\figurename~\ref{fig:sys} presents the design. \sys\ isolates variable-shaped
prefill work (Section~\ref{sec:prefill}), pads all decode iterations to one captured
shape (Section~\ref{sec:decode}), and aligns sampling with the same fixed request
layout (Section~\ref{sec:sampling}).

\subsection{Determinism Contract}

Let $Y(x,s;\mathcal{C},p,a)$ denote the output token sequence for prompt $x$
and per-request sampling seed $s$, where $\mathcal{C}$ is the set of
co-running requests, $p$ is the request's batch position, and $a$ is the
request arrival order. Given the same model weights, sampling parameters,
hardware, software stack, and tensor-parallel configuration, \sys\ provides
\[
  Y(x,s;\mathcal{C},p,a)
  =
  Y(x,s;\mathcal{C}',p',a')
\]
for arbitrary valid co-runners, positions, and arrival orders.

\sys\ establishes this contract by making each target request observe the same
request-local inputs and arithmetic order. Isolated prefill removes other
requests from the prompt computation. Fixed-shape decode makes the launch
shape independent of the number of active requests. Position-invariant
kernels then tolerate slot changes within that shape. \sys\ retains a fixed
tree for collective reductions and disables decode-time attention Split-KV,
the two components in \tablename~\ref{tab:kernel-invariance} that are not
position-invariant. Finally, request-bound RNG state
makes seeded multinomial sampling independent of the batch slot.

The contract does not claim reproducibility across hardware architectures,
kernel versions, numerical formats, or tensor-parallel degrees. Those changes
can alter the arithmetic itself and require orthogonal techniques such as
cross-configuration invariant kernels~\cite{zhang2026deterministic}.

\subsection{Isolated Prefill}\label{sec:prefill}

A simple way to standardize prefill would pad every prompt to one global
length. This design is inefficient because prompt lengths vary by orders of
magnitude. A large target length wastes computation on short requests, while
a small target length forces long prompts through many chunked-prefill
iterations. \sys\ instead makes each prefill request its own batch. A prompt
can retain its natural length, but its execution shape no longer depends on
the lengths or arrival times of other requests.

Isolation also preserves an important prefill optimization. Split-KV may
choose its partitioning from the prompt length and kernel configuration.
Because the same request always enters prefill alone, its prompt and chunk
shapes remain unchanged across executions. \sys\ can therefore retain
prefill-time Split-KV without requiring the attention kernel to be
position-invariant. This choice avoids the substantial loss that would result
from disabling Split-KV for long prompts.

\sys\ implements isolation as a scheduling wrapper around the original vLLM
scheduler. Algorithms~\ref{alg:1} and~\ref{alg:2} show the relevant logic.
The scheduler maintains a \emph{running} queue containing requests that have
started execution and a \emph{waiting} queue containing requests that have
not started prefill. A request at the head of \emph{running} may still require
prefill when a long prompt has been divided into fixed-size chunks. In that
case, Algorithm~\ref{alg:1} selects the partial request before admitting new
work, preserving its chunk sequence (lines 2-6).
\sys\ starts a waiting request only when the number of requests currently being processed has not exceeded the maximum limit (lines 7-9) and the KV-cache capacity can accommodate
its full configured sequence (line 10). This rule prevents a request from being
preempted or recomputed because a later decode step exhausts capacity. If no
request needs prefill, the wrapper invokes the original scheduler and allows a
normal decode iteration.

\begin{algorithm}
  \caption{Finding the request for isolated prefill}
  \label{alg:1}
  \begin{algorithmic}[1]
    \INPUT Request queues $running$ and $waiting$
    \OUTPUT A request requiring prefill, or \textbf{none}
    \PARAMETER The maximum number of running requests $max\_num\_reqs$.
    \Function{find\_prefill\_request}{}
      \If{$running \ne \emptyset$}
        \State head $\leftarrow running[0]$
        \If{(head.num\_computed\_tokens $<$ \\ $\qquad\qquad\quad$head.num\_prompt\_tokens)}
          \State \Return head \Comment{Continue chunked prefill}
        \EndIf
      \EndIf
      \If{($waiting \ne \emptyset$ \&\\ $\qquad\quad$\textsc{length}($running$) $<$ $max\_num\_reqs$)}
        \State head $\leftarrow waiting$.\textsc{peek\_request}()
        \If{\textsc{can\_fit\_full\_sequence}(head)}
          \State \Return head
        \EndIf
      \EndIf
      \State \Return \textbf{none}
    \EndFunction
  \end{algorithmic}
\end{algorithm}

When Algorithm~\ref{alg:1} selects a request, Algorithm~\ref{alg:2} temporarily
isolates it without replacing the original scheduler. The algorithm first
calls Algorithm~\ref{alg:1} to find a request requiring prefill (line 2). If
no such request exists, it immediately invokes the original scheduler, which
produces a normal decode iteration (lines 3--4). Otherwise, it copies both
request queues so that the requests hidden during isolated prefill can be
restored afterward (lines 5--6). It then clears the waiting queue to prevent
the original scheduler from admitting another request into the prefill batch
(line 7).

Lines 8--14 expose only the selected request to the original scheduler. If the
request has already started chunked prefill, it resides in $running$; the
algorithm replaces $running$ with a singleton queue and removes the request
from the stashed copy (lines 8--10). Otherwise, the request has not started
prefill. The algorithm clears $running$, inserts the selected request into
$waiting$, and removes it from the stashed waiting queue to avoid restoring a
duplicate entry (lines 11--14). The call to \textsc{schedule\_origin} then
produces an ordinary scheduling result for this single request (line 15).
Finally, the algorithm appends the stashed running requests and reinserts the
stashed waiting requests in their original order (lines 16--18), then returns
the isolated-prefill result (line 19). The model runner therefore executes an
ordinary prefill iteration with batch size one; isolation requires no separate
model-execution path.

\begin{algorithm}
  \caption{Scheduling isolated prefill}
  \label{alg:2}
  \begin{algorithmic}[1]
    \INPUT Request queues $running$ and $waiting$
    \OUTPUT Scheduling result $output$
    \Function{schedule}{}
      \State det $\leftarrow$ \textsc{find\_prefill\_request}()
      \If{det $==$ \textbf{none}}
        \State \Return \textsc{schedule\_origin}()
      \EndIf
      \State stashed\_running $\leftarrow$ \textsc{copy}($running$)
      \State stashed\_waiting $\leftarrow$ \textsc{copy}($waiting$)
      \State $waiting$.\textsc{clear}()
      \If{det in $running$}
        \State $running \leftarrow $[det]
        \State stashed\_running.\textsc{remove}(det)
      \Else
        \State $running$.\textsc{clear}()
        \State $waiting$.\textsc{add\_request}(det)
        \State stashed\_waiting.\textsc{remove}(det)
      \EndIf
      \State $output \leftarrow$ \textsc{schedule\_origin}()
      \State $running \leftarrow running + stashed\_running$
      \For{request in stashed\_waiting}
        \State $waiting$.\textsc{add\_request}(request)
      \EndFor
      \State \Return $output$
    \EndFunction
  \end{algorithmic}
\end{algorithm}

\subsection{Fixed-Shape Batched Decode}\label{sec:decode}

Each active request contributes exactly one query token to a decode iteration.
\sys\ therefore chooses the engine's maximum concurrency,
$max\_num\_reqs$, as the fixed decode shape. High-throughput offline workloads
usually keep the running queue near this limit, so most slots perform useful
work. When fewer requests are active, masked padding tokens occupy the unused
slots.

\sys\ realizes fixed-shape decode through the existing CUDA Graph interface, without modifying the model runner or CUDA Graph implementation. Our
configuration uses:
\begin{verbatim}
  -cc.cudagraph_mode=full_decode_only \
  --cudagraph-capture-sizes 256 \
  --max-num-reqs 256
\end{verbatim}
The first option restricts graph execution to decode. The second captures one
graph for 256 query tokens, and the third limits the engine to 256 concurrent
sequences. Since isolated prefill prevents mixed prefill-decode iterations,
every decode iteration contains at most 256 decode tokens. The engine
pads that iteration to the only captured size, making its launch shape exactly
256. The capture size is configurable; we use 256 to match the maximum concurrency in our evaluation.

This construction reuses the engine's existing graph padding and masking
logic; \sys\ does not add a second model-execution path. It also amortizes the
cost of padding through graph replay. Padding can be expensive when only a few
requests remain, but it has little cost while the engine is highly utilized.
This tradeoff motivates \sys's focus on batched evaluation and RL rollout
workloads rather than low-concurrency interactive serving.

The fixed query count alone is insufficient for attention because requests
have different KV lengths. A Split-KV implementation can vary the number and
placement of KV partitions according to the ragged batch. \sys\ disables
Split-KV during decode so that each request follows a request-local reduction
order independent of its neighbors. Other model kernels use the same launch
shape and are position-invariant, allowing requests to move among the 256
slots as earlier requests finish.

\subsection{Shape-Aligned Sampling}\label{sec:sampling}

The model runner produces logits, but the sampler converts those logits into
the next token. Sampling executes outside the captured CUDA Graph in our
stack, so graph padding does not automatically standardize its input. During
prefill, the engine selects only the final prompt token's hidden state; isolated
prefill therefore presents one sampling row. During decode, \sys\ pads the
sampler input to $max\_num\_reqs$, matching the fixed model-output shape.

\sys\ supports seeded multinomial sampling by binding RNG state to a request
rather than to a batch slot. Padding entries never consume a real request's
RNG state. Removing a completed request or moving a live request to another
slot also leaves every other request's RNG progression unchanged. The target
request consequently receives the same random variate at each generation
step. Given the identical logits produced by the model path, the same
per-request random variate selects the same token and completes the
determinism contract.

\subsection{Determinism Argument}

We now show that the mechanisms above satisfy the contract stated at the
beginning of this section. The argument assumes the fixed hardware, software,
model, and tensor-parallel configuration specified by the contract. It also
uses the verified properties in \tablename~\ref{tab:kernel-invariance}: every kernel exercised by the evaluated model configurations has been verified to be position-invariant or batch-invariant~\cite{gond2026llm42}, and
each request's model computation is isolated from the values of other
requests. The latter property follows from request-local attention masks, KV
caches, and token-wise model operators. Co-runners may affect the execution
shape, but cannot participate in the target request's mathematical
computation.

Let $z_t^r$ denote the state of request $r$ before generation step $t$. It
contains the prompt and generated prefix, the request's KV cache, and its sampling parameters. Consider two executions, $A$ and $B$, with the same prompt $x$ and
request seed $s$, but arbitrary co-runners, batch positions, and arrival
orders. We prove by induction that
\[
    z_t^{r,A}=z_t^{r,B}
    \quad\text{and}\quad
    y_{<t}^{r,A}=y_{<t}^{r,B}
    \qquad \forall t,
\]
where $y_{<t}^r$ is the generated token prefix before step $t$.

\paragraph{Base case}
Isolated prefill executes request $r$ without co-runners. Because both
executions have the same prompt and configuration, they present the same
prefill shapes and values to every kernel. Each kernel therefore follows the
same tiling and reduction order, producing identical logits and KV-cache
states:
\[
    z_0^{r,A}=z_0^{r,B}.
\]

The request-bound RNG is initialized from the same seed $s$, so its initial
state is also identical.

\paragraph{Inductive step}
Assume $z_t^{r,A}=z_t^{r,B}$. \sys\ pads both decode batches to the same
captured size $N=max\_num\_reqs$. Request $r$ may occupy positions $p_A$ and
$p_B$, and the remaining $N-1$ positions may contain different requests or
padding. Let $X_t^A$ and $X_t^B$ denote the two padded decode batches at step
$t$, including their request-local tokens, KV-cache metadata, attention
masks, and padding entries. Let $D_N$ denote one fixed-shape model step and
$\pi_p$ select the output at position $p$. Request isolation and position
invariance give
\[
  d(z_t^{r,A})
  =
  \pi_{p_A}D_N(X_t^A)
  =
  \pi_{p_B}D_N(X_t^B)
  =
  d(z_t^{r,B}),
\]
where $d$ is the request-local decode function. The equality also covers
collectives and attention because \sys\ uses fixed-tree reductions and
disables decode-time Split-KV. Thus, the two executions produce identical
logits $\ell_t^{r,A}=\ell_t^{r,B}$ and identical updated KV-cache values.

Let $R(s,t)$ be the random variate consumed by request $r$ at step $t$, and
let $S$ be the deterministic sampling function. Since RNG state is bound to
the request and padding does not consume it,
\[
  y_t^{r,A}
  = S(\ell_t^{r,A},R(s,t))
  = S(\ell_t^{r,B},R(s,t))
  = y_t^{r,B}.
\]

Appending the same token to identical prefixes and KV-cache states yields
$z_{t+1}^{r,A}=z_{t+1}^{r,B}$, completing the induction.

Therefore, the complete output sequence is identical:
\[
  Y(x,s;\mathcal{C},p,a)
  =
  Y(x,s;\mathcal{C}',p',a').
\]

Arrival order and co-runners can change when a request executes, but under
\sys\ they cannot change its sequence of model states, random variates, or
sampled tokens.

\section{Evaluation}

We evaluate whether \sys\ (1) preserves token-sequence determinism, (2)
recovers the throughput lost by batch-invariant inference, and (3) remains
effective across models, workloads, and concurrency levels.

\subsection{Experimental Setup}

\paragraph{Models}
We select three large open models with different architectures:
Qwen3-235B-A22B~\cite{qwen3technicalreport},
DeepSeek-V3~\cite{deepseekai2024deepseekv3technicalreport}, and
Hy3~\cite{tencent2026hy3}.

\begin{figure}
    \centering
    \includegraphics[width=0.5\linewidth]{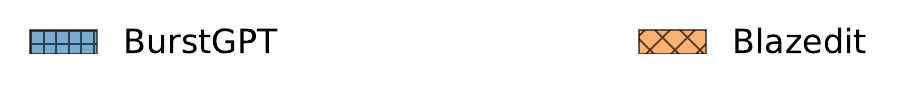} \\
    \begin{subfigure}[t]{0.48\linewidth}
        \includegraphics[width=\linewidth]{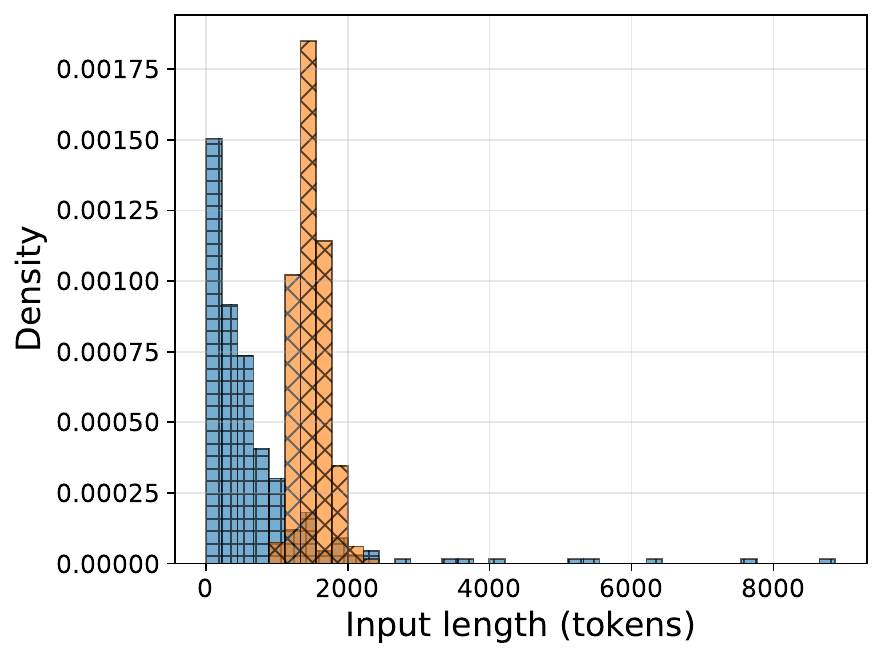}
        \caption{Input Length}
    \end{subfigure}
    \begin{subfigure}[t]{0.48\linewidth}
        \includegraphics[width=\linewidth]{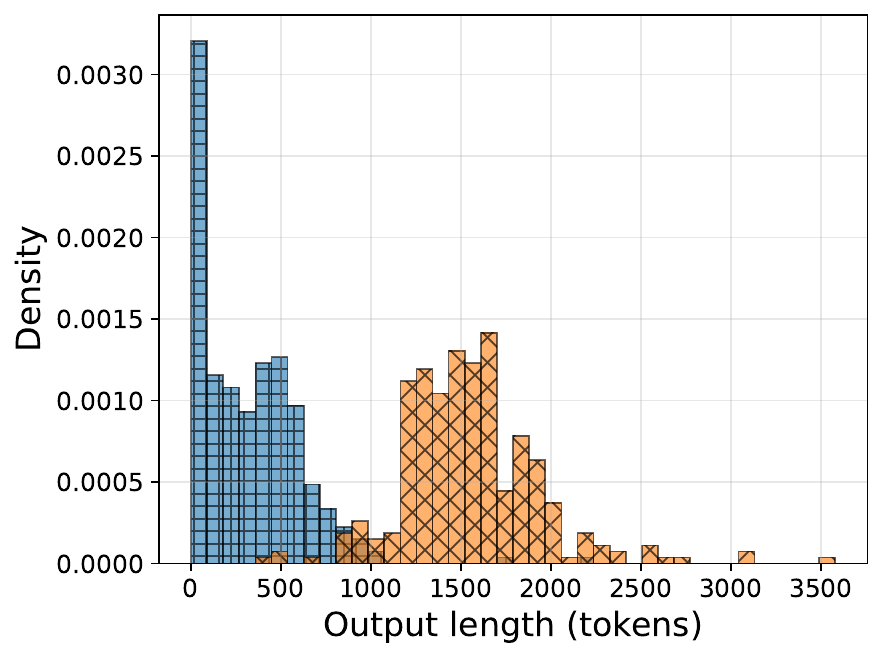}
        \caption{Output Length}
    \end{subfigure}
    \caption{Input and output length distribution.}
    \label{fig:data}
\end{figure}

\paragraph{Workloads}
We evaluate three workload classes and sample 300 requests from each workload.
BurstGPT contains traces collected from regional Azure OpenAI services and
captures the burstiness and length variation of production
traffic~\cite{wang2025burstgpt}. Its length distribution is long-tailed, with
a small number of prompts and responses exceeding 8\,K tokens, as shown in \figurename~\ref{fig:data}.
Blazedit is a code-editing workload characterized by long inputs and long
outputs~\cite{daita2025blazedit}. We also construct synthetic workloads with
fixed input and output lengths to isolate how each length affects \sys.

\paragraph{Baselines}
We implement \sys\ in vLLM 0.25.0~\cite{kwon2023efficient}. We compare three
configurations. Non-determinism (\emph{STD}) is the default vLLM execution path, which does not enforce deterministic outputs. It
represents the performance ceiling of a system without determinism
constraints. Batch-invariant (\emph{BI}) enables vLLM's batch-invariant execution. \sys\ 
uses isolated prefill, fixed-shape decode, shape-aligned sampling, and the
fixed-tree collective and decode-time Split-KV policy described in
Section~\ref{sec:decode}. All configurations use the same model,
tensor-parallel degree, memory limits, and request set.

\paragraph{Metrics}
Our primary metric is end-to-end output token throughput because model evaluation and
RL rollout process a finite batch of requests and seek to minimize its total
completion time. For a collection of tasks, we report aggregate throughput:
\[
  \text{Aggregate Throughput}
  =
  \frac{\sum_{i \in \mathrm{Task}}
    \text{Throughput}_i \times \text{Time}_i}
  {\sum_{i \in \mathrm{Task}} \text{Time}_i}.
\]

This time-weighted average gives greater weight to tasks that occupy the
system for longer. An unweighted arithmetic mean would assign the same
importance to a short task and a long-running task, even though optimizing the
latter has a larger effect on the total time required to finish the complete
workload. We also report time-to-first-token
(TTFT) and time-per-output-token (TPOT) to expose the latency effects of
isolated prefill and fixed-shape decode.

\subsection{Determinism Validation}

We verify the determinism of our method using vLLM's batch-invariance test harness~\cite{vllm2026test}. It compares isolated and
batched execution under seeded multinomial sampling. For each evaluated model,
the test generates 32 prompts and executes each prompt first at batch size
one and then as part of a batch of 32. It uses temperature 0.6,
seed 42, and generates up to 8 output tokens per prompt. The comparison
checks the complete generated token sequence and the sampled-token log
probability at every generation step.

\begin{figure}
    \centering
    \includegraphics[width=\linewidth]{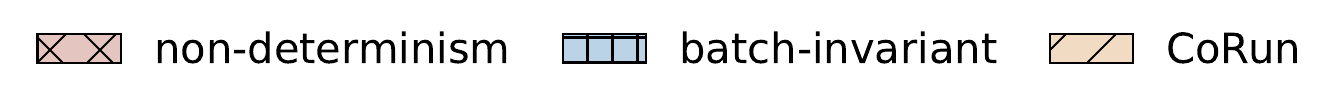}
    \begin{subfigure}[t]{\linewidth}
        \includegraphics[width=\linewidth]{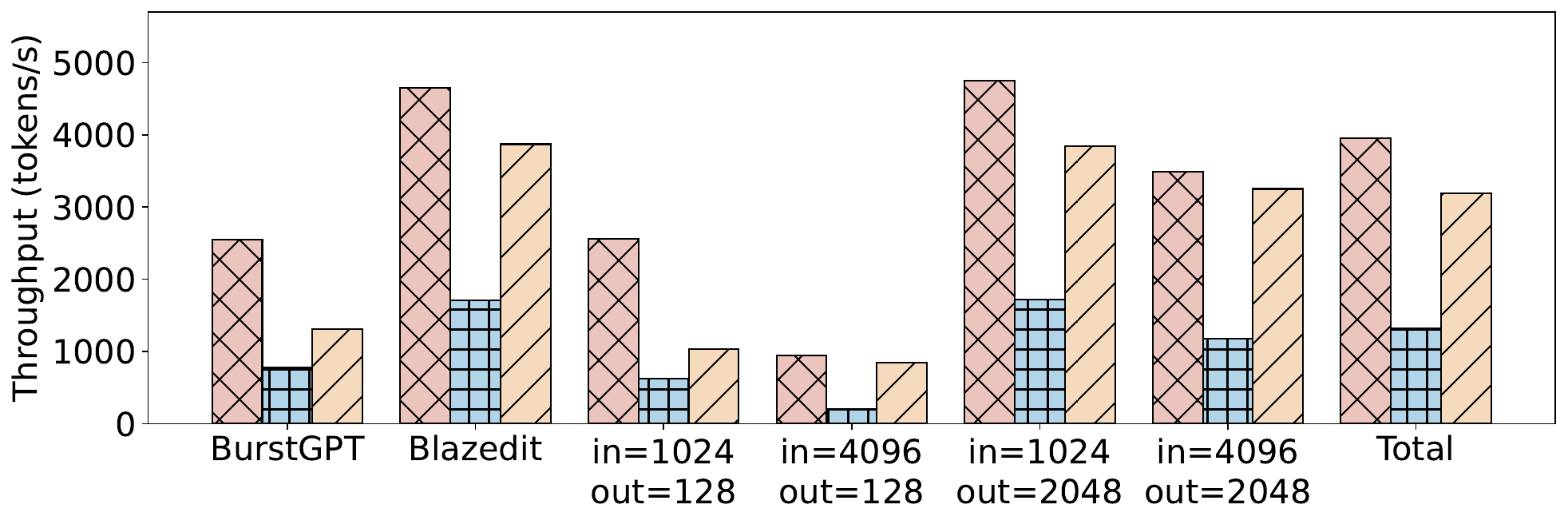}
        \caption{Qwen3-235B-A22B}
    \end{subfigure}
    \begin{subfigure}[t]{\linewidth}
        \includegraphics[width=\linewidth]{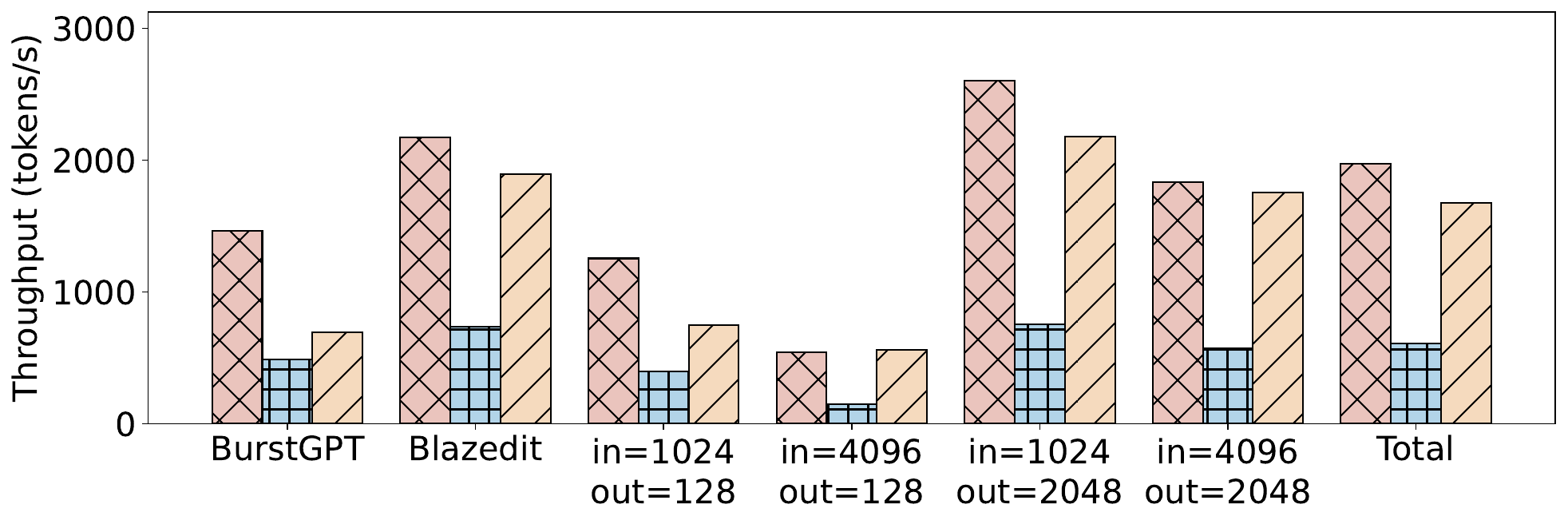}
        \caption{DeepSeek-V3}
    \end{subfigure}
    \begin{subfigure}[t]{\linewidth}
        \includegraphics[width=\linewidth]{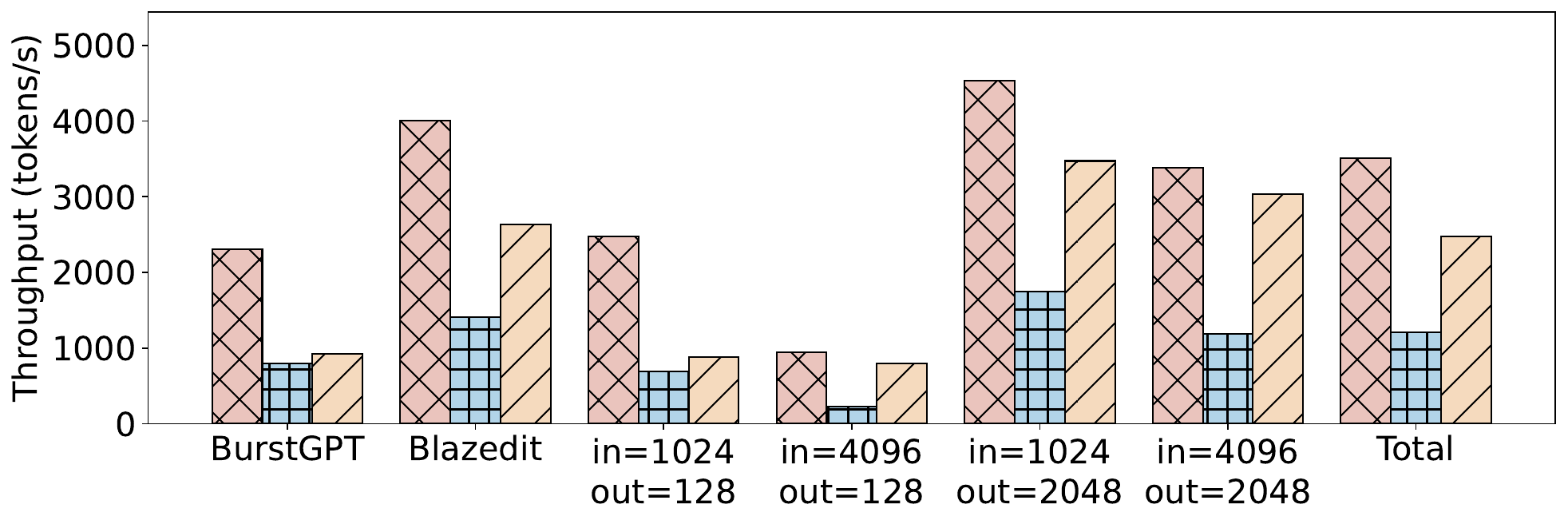}
        \caption{Hy3}
    \end{subfigure}
    \caption{Throughput evaluation on diverse datasets and models. ``in=$m$, out=$n$'' refers to a dataset with an input length of $m$ and an output length of $n$.}
    \label{fig:throughput}
\end{figure}
All 32 prompts pass for each of the three models: \sys\ returns identical token
sequences and exactly matching sampled-token log probabilities between
isolated and batched execution. This test changes both the co-running request
set and the batch position for all prompts. Together with the
request-independent execution shapes and request-bound RNG state established
in Section~\ref{sec:sampling}, the result validates the batch-composition and
position components of \sys's determinism contract for the evaluated stack.

\subsection{End-to-End Throughput}\label{sec:throughput}

\figurename~\ref{fig:throughput} reports throughput across the three models and
six workload configurations. \sys\ improves throughput by 15--324\% over BI.
Its aggregate throughput more than doubles for every evaluated model. These
results show that replacing broad batch invariance with fixed-shape scheduling
recovers a large fraction of the performance available to STD.

The gain varies with workload shape. \sys's smallest improvements occur on
BurstGPT. Its long tail leaves only one request active near the end of the
run, so \sys\ continues to replay the fixed-size decode graph while most slots
contain padding. The synthetic workloads and Blazedit maintain useful
concurrency for longer and amortize that padding across more active requests.
They also expose the high kernel cost of BI, producing larger relative gains.

Notably, on DeepSeek-V3 with input length 4096 and output length 128, \sys\ even exceeds
STD. We found that for this workload, the standard scheduler frequently combines prefill and decode requests. DeepSeek-V3 uses the \textsc{FLASH\_ATTN\_MLA} attention
backend, which cannot use a CUDA Graph for attention when the batch mixes
prefill and decode. \sys\ separates the two stages, allowing every decode
iteration to replay the captured graph. This graph benefit outweighs \sys's
padding and isolation overhead in this configuration. Models using the
\textsc{FLASH\_ATTN} backend can retain graph execution for mixed batches, so
their STD configurations do not receive the same penalty.

\begin{figure}
    \centering
    \includegraphics[width=\linewidth]{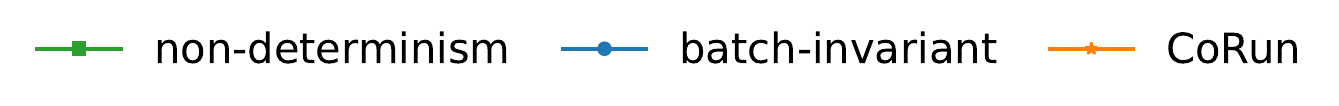}
    \begin{subfigure}[t]{\linewidth}
        \includegraphics[width=\linewidth]{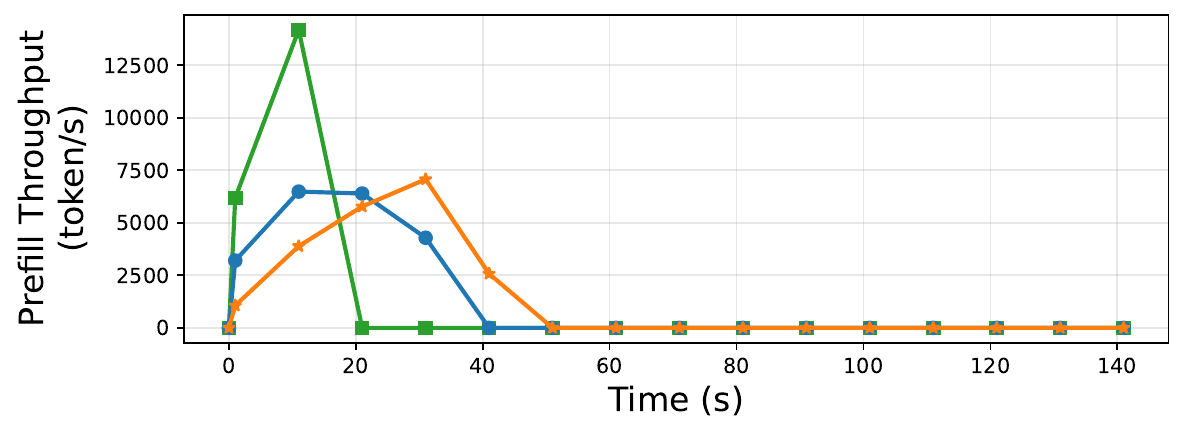}
        \caption{Prefill throughput over time}
        \label{fig:prefill}
    \end{subfigure}
    \begin{subfigure}[t]{\linewidth}
        \includegraphics[width=\linewidth]{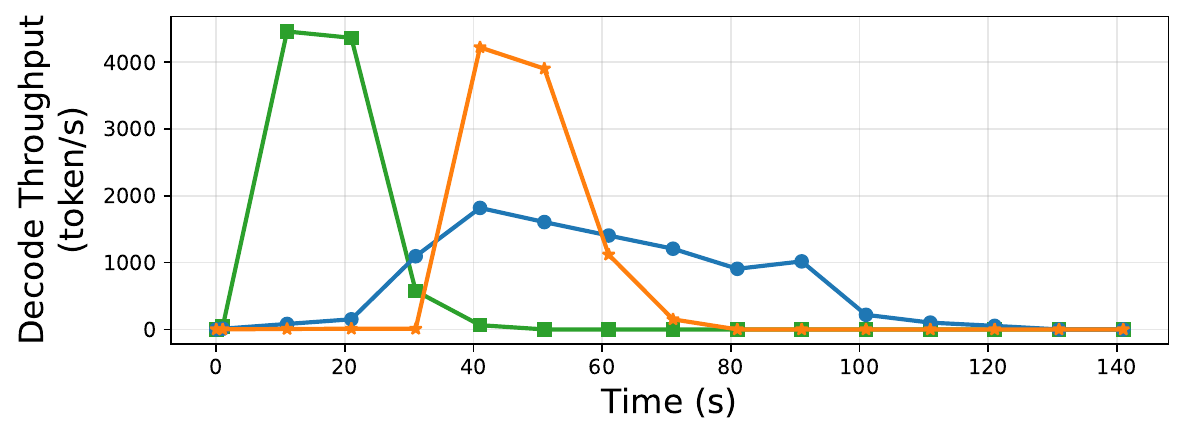}
        \caption{Decode throughput over time}
        \label{fig:decode}
    \end{subfigure}
    \caption{Prefill and decode throughput over time using the Qwen3-235B-A22B model and the BurstGPT dataset.}
    \label{fig:details}
\end{figure}

\paragraph{Throughput over time}
\figurename~\ref{fig:details} separates prefill and decode throughput over
time on BurstGPT. At time zero, 256 requests arrive, so
all three systems initially devote most of their capacity to prefill and shift
toward decode later in the run. The client submits a new request whenever one
request completes, until it has submitted all 300 requests.

STD progresses through both stages fastest. It completes all prefill work in 20 seconds and all decode work in 37 seconds. BI
has substantially lower throughput in both stages, requiring approximately
40 seconds to finish prefill and 124 seconds to finish decode. \sys\ instead
front-loads isolated prefill: it performs no decode work during the first
30 seconds while processing prompts one at a time. This serialized prefill is
its primary source of overhead. Once \sys\ enters its decode-intensive phase,
however, its decode throughput approaches that of STD and reaches 83\% of
STD's peak. The large decode-throughput improvement over BI accounts for most
of \sys's end-to-end gain and makes it particularly effective for
decode-intensive workloads such as RL rollout.

\subsection{Latency Analysis}

\begin{figure}
    \centering
    \includegraphics[width=\linewidth]{figs/legend_cdf.pdf} \\
    \begin{subfigure}[t]{\linewidth}
        \includegraphics[width=\linewidth]{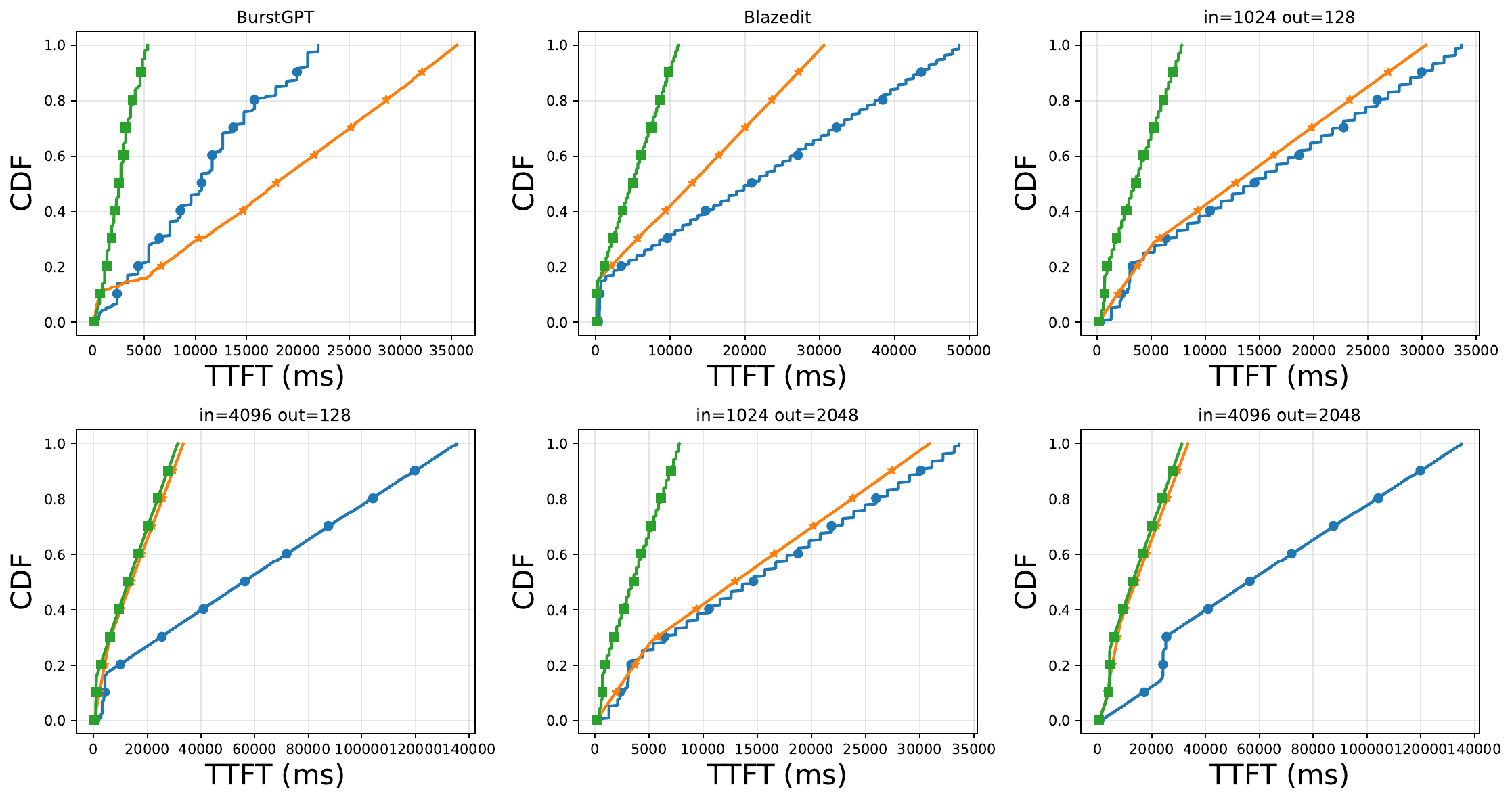}
        \caption{Time to first token}
    \end{subfigure}
    \begin{subfigure}[t]{\linewidth}
        \includegraphics[width=\linewidth]{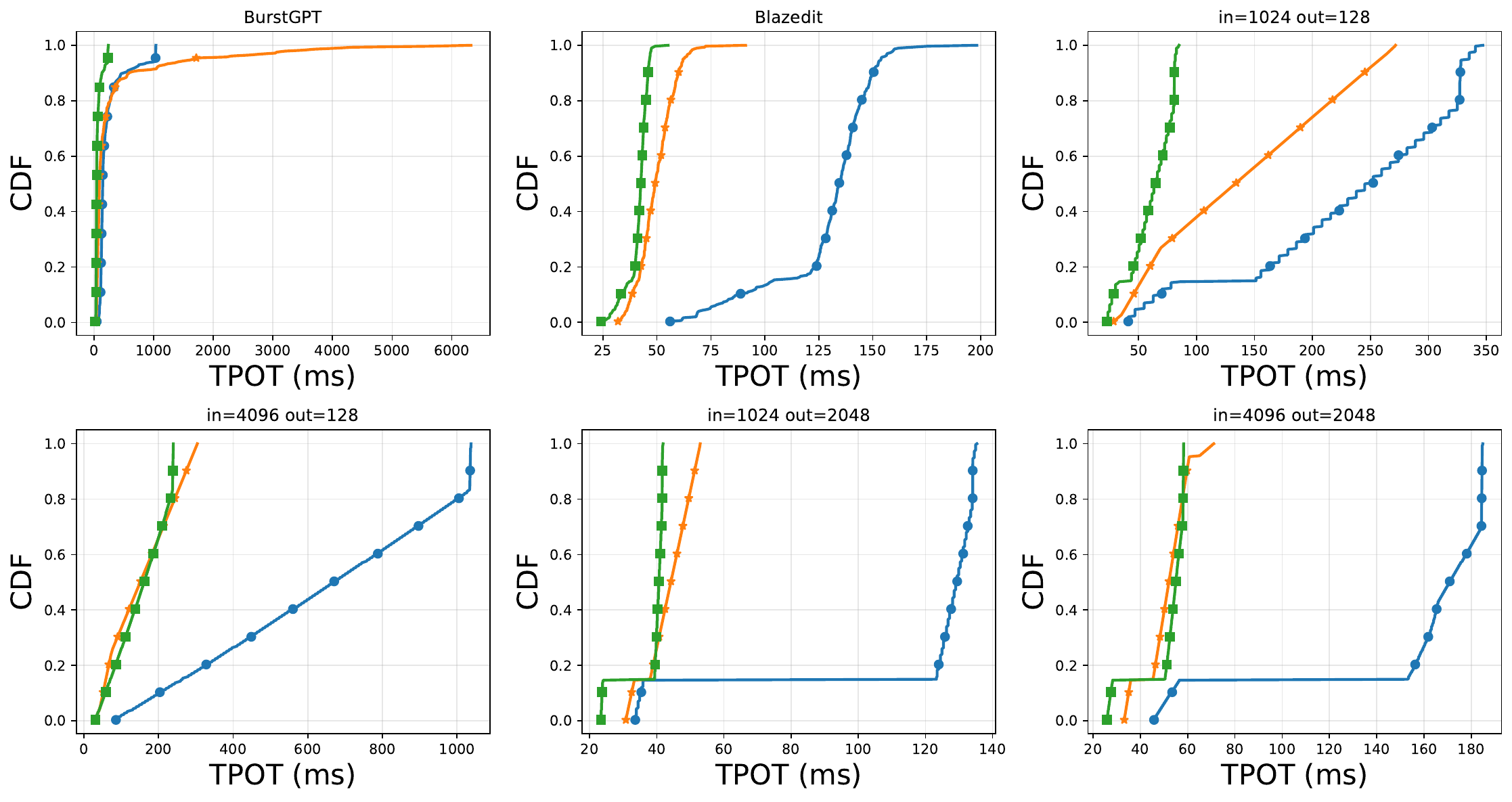}
        \caption{Time per output token}
    \end{subfigure}
    \caption{Latency analysis on Hy3.}
    \label{fig:latency}
\end{figure}

\figurename~\ref{fig:latency} shows the TTFT and TPOT distributions for Hy3. \sys\
generally lies between STD and BI. STD can batch or chunk multiple prefills and
therefore provides the lowest latency, while BI pays the latency of slower
invariant kernels. \sys\ retains fast kernels but serializes requests' prefill work. Across the six workloads, \sys\ reduces mean TTFT by
51.8\% and mean TPOT by 48.6\% relative to BI.

Isolated prefill produces a staircase in TTFT: requests wait while earlier
prompts complete their prefill stages, and then advance as each isolated
prompt leaves the queue. The effect is most visible on BurstGPT. Requests that
arrive before a long prompt see short waits, while requests behind that prompt
inherit its long prefill time. Decode padding similarly creates a TPOT tail
when BurstGPT's active set drains. Despite these scheduling costs, \sys\
retains lower TTFT and TPOT than BI for most of the distribution because it
avoids the latency of broadly batch-invariant kernels.

\subsection{Scalability with Concurrency}

\begin{figure}
    \centering
    \begin{subfigure}[t]{\linewidth}
        \includegraphics[width=\textwidth]{figs/legend_cdf.pdf}
    \end{subfigure}
    \begin{minipage}{0.49\columnwidth}
        \centering
        \includegraphics[width=\textwidth]{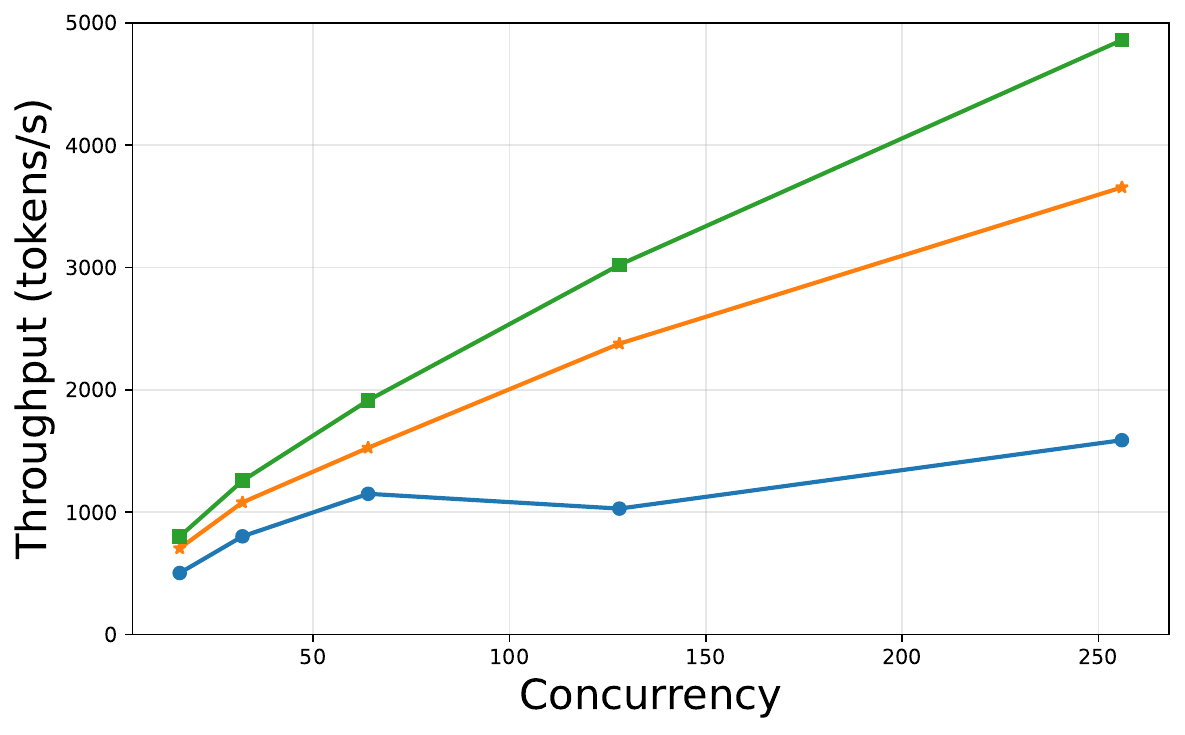}
        \caption{Scalability evaluation under different concurrencies.}
        \label{fig:concurrency}
    \end{minipage}
    \hfill
    \begin{minipage}{0.49\columnwidth}
        \centering
        \includegraphics[width=\textwidth]{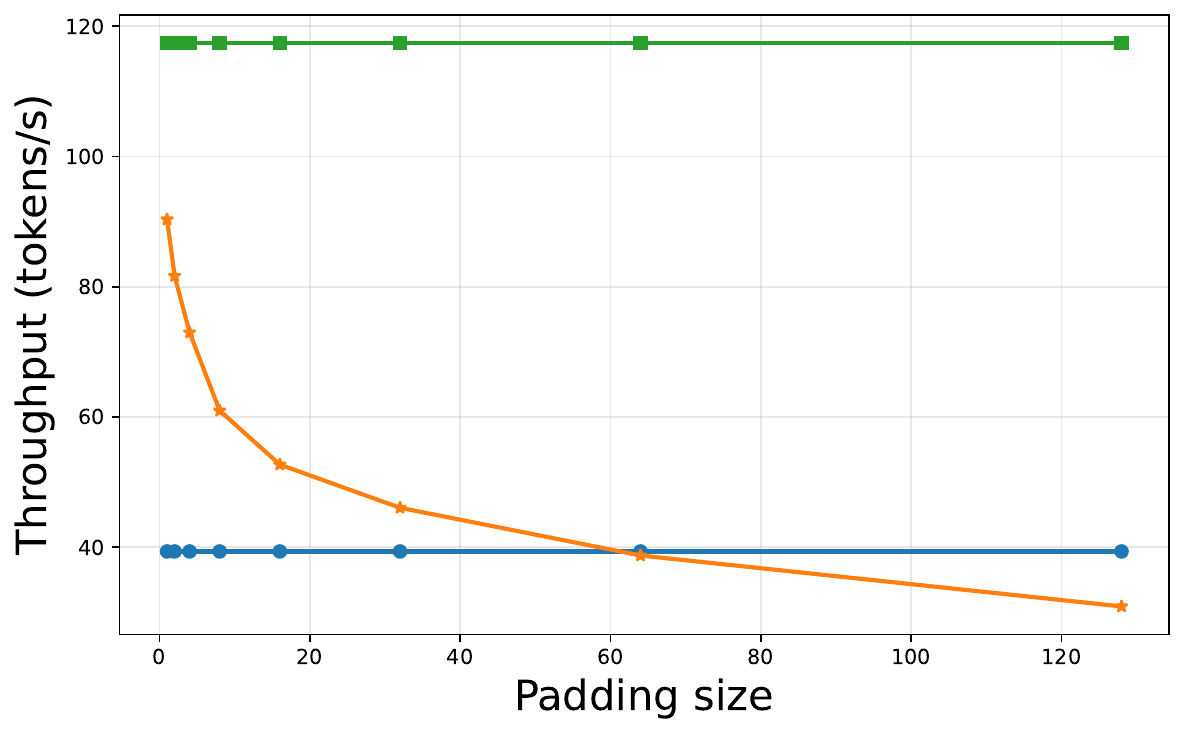}
        \caption{The impact of padding in the decode phase.}
        \label{fig:padding}
    \end{minipage}
\end{figure}

\figurename~\ref{fig:concurrency} evaluates Hy3 on Blazedit while varying the
maximum number of concurrent requests. \sys's throughput rises with
concurrency because more fixed decode slots perform useful work and isolated
prefills are amortized over a longer period of batched decoding. BI improves
more slowly because its invariant kernels remain the bottleneck as the batch
grows. \sys\ therefore benefits from the high concurrency available in offline
evaluation and rollout workloads.

\subsection{Ablation Study}

\sys\ differs from STD in three performance-relevant ways: it pads decode to a
fixed shape, disables attention Split-KV during decode, and isolates prefill.
We isolate these costs to identify when the scheduling approach is most
effective.

\paragraph{Decode padding}
\figurename~\ref{fig:padding} measures the worst case for fixed-shape decode. Only
one request is active while we vary \sys's padding size; STD and BI do not
pad and therefore remain constant. \sys\ throughput decreases approximately
inversely with the padding size because nearly all added tokens are redundant.
When the padding size exceeds 64 in this single-request experiment, \sys\
falls below BI. This result defines the system's intended operating region:
\sys\ targets high-concurrency offline inference, where most graph slots are
occupied, rather than single-request serving.

\begin{figure}
    \centering
    \includegraphics[width=\linewidth]{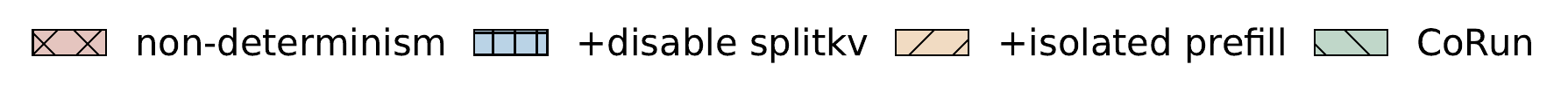}
    \includegraphics[width=\linewidth]{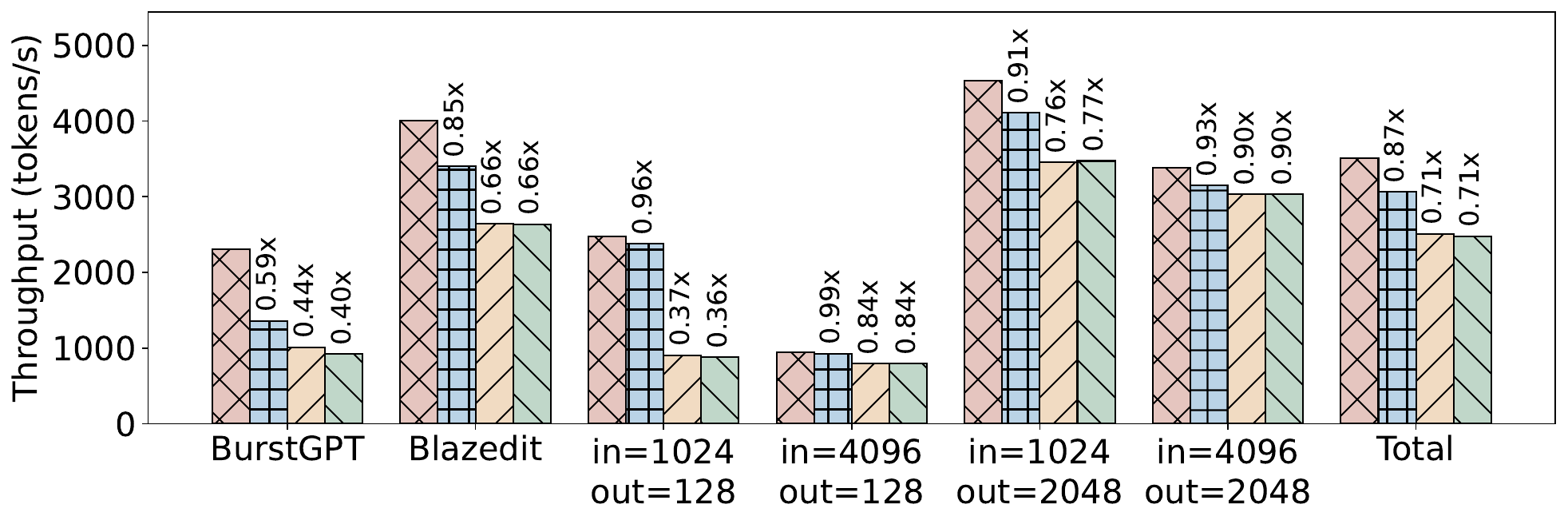}
    \caption{Ablation study.}
    \label{fig:ablation}
\end{figure}

\paragraph{Split-KV and isolated prefill}
\figurename~\ref{fig:ablation} starts from STD and adds the remaining constraints.
\emph{+disable splitkv} disables Split-KV only for decode attention; isolated
prefill is disabled. \emph{+isolated prefill} then separates
prefill from all concurrent work and completes the \sys\ scheduling policy.

Disabling Split-KV costs 15--41\% when prompt lengths vary, because the ragged
batch offers uneven attention work that benefits from additional KV
partitions. The loss falls to 1--9\% when input lengths are similar. Isolated
prefill accounts for the larger share of \sys's gap to STD because
it gives up batching across prompts and prevents prefill-decode mixing.

The isolation cost decreases as inputs grow. Long prompts reduce the number of
prefill requests that can profitably share an iteration, limiting the benefit
of batching them in STD. The cost also decreases as outputs grow because
decode occupies a larger fraction of end-to-end execution. \sys\ is therefore
particularly well suited to workloads with long inputs, long outputs, or both,
including the code-editing and RL rollout workloads that motivate deterministic
offline inference.

\section{Related Work}

\paragraph{Numerical nondeterminism in LLM inference}
Empirical studies show that nominally deterministic API settings can still
produce different text and task scores. At{\i}l et al.\ observe output
variation across all tested hosted models and report a gap of up to 70\%
between the best and worst task outcomes available across repeated
runs~\cite{atil2025non}. Yuan et al.\ trace such divergence to numerical
sources within LLM inference, including precision and batch-dependent
computation, and show that higher-precision accumulation reduces but does not
eliminate the problem~\cite{yuan2025fp32}. These effects matter beyond
evaluation~\cite{rainbird2025}. Training-inference mismatch can cause rollout tokens and
training-time log probabilities to disagree, independently degrading LLM
reinforcement learning~\cite{feng2025offpolicy,zhong2026diagnosing}.

\paragraph{Deterministic LLM inference}
Thinking Machines, SGLang, and vLLM obtain deterministic serving by replacing
batch-dependent reductions with batch-invariant kernels and deterministic
collectives~\cite{he2025nondeterminism,sglang2025det,vllm2026det}. DeepSeek-V4
similarly uses deterministic execution to support debugging, stability
analysis, and reproducible post-training behavior~\cite{deepseekai2026deepseekv4}.
TBIK extends invariance across tensor-parallel degrees by mapping reductions
onto a unified hierarchical binary tree~\cite{zhang2026deterministic}. \sys\
addresses a different axis: it keeps the model and tensor-parallel
configuration fixed, but prevents dynamic request batching from changing the
execution of a target request.

LLM-42 avoids running every token through batch-invariant kernels. It generates
tokens on a fast dynamic path, periodically verifies them with a deterministic
fixed-shape path, and rolls back when verification
fails~\cite{gond2026llm42}. MarginGate reduces this verification load by
triggering deterministic verification only when the decision margin indicates
that a numerical perturbation could change the selected token~\cite{chu2026margingate}.
These methods preserve flexible batching through speculation and verification.
\sys\ instead constrains prefill and decode shapes before execution. It does
not generate unverified tokens, invoke a separate verifier, or roll back
requests. This design targets workloads in which every request requires
determinism and sufficient concurrency is available to amortize fixed-shape
decode.

\paragraph{LLM serving and scheduling}
Orca introduced iteration-level scheduling to admit and retire requests
between decoding steps~\cite{yu2022orca}. vLLM combines continuous batching
with paged KV-cache management to sustain larger dynamic
batches~\cite{kwon2023efficient}, while SGLang adds a structured runtime and
cache-aware execution for complex model programs~\cite{zheng2024sglang}.
Sarathi-Serve uses chunked prefill to bound generation stalls and enable
stall-free batching of prefill and decode~\cite{agrawal2024sarathi}. Other
systems disaggregate or independently schedule the two stages to meet
throughput and latency objectives~\cite{patel2024, qin2025mooncake, zhong2024}. \sys\ deliberately restricts this
scheduling freedom: it serializes prefill across requests and standardizes
decode shape to make optimized position-invariant kernels deterministic.

\section{Limitations and Design Tradeoffs}

As discussed in the previous section, LLM-42 uses verified speculation to retain
dynamic batching~\cite{gond2026llm42}. This design complements \sys\ in two
scenarios where fixed-shape scheduling is less suitable.

\subsection{Sparse Deterministic Traffic}

LLM-42 can apply verification only to requests that require deterministic
outputs, while all other requests remain on the unmodified fast path. Its
overhead therefore scales with the fraction of deterministic traffic, making
it attractive when that fraction is small.

\begin{figure}
    \centering
    \begin{subfigure}[t]{0.48\linewidth}
        \includegraphics[width=\linewidth]{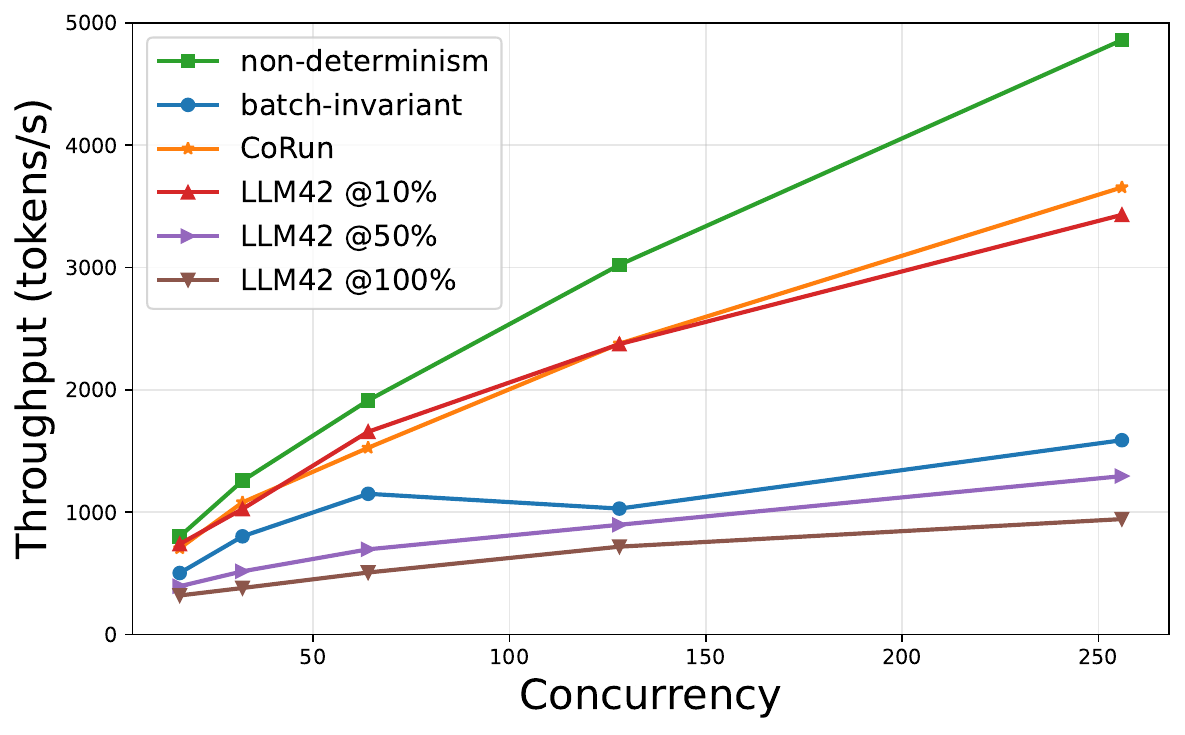}
        \caption{Throughput under mixed deterministic traffic}
        \label{fig:llm42a}
    \end{subfigure}
    \begin{subfigure}[t]{0.48\linewidth}
        \includegraphics[width=\linewidth]{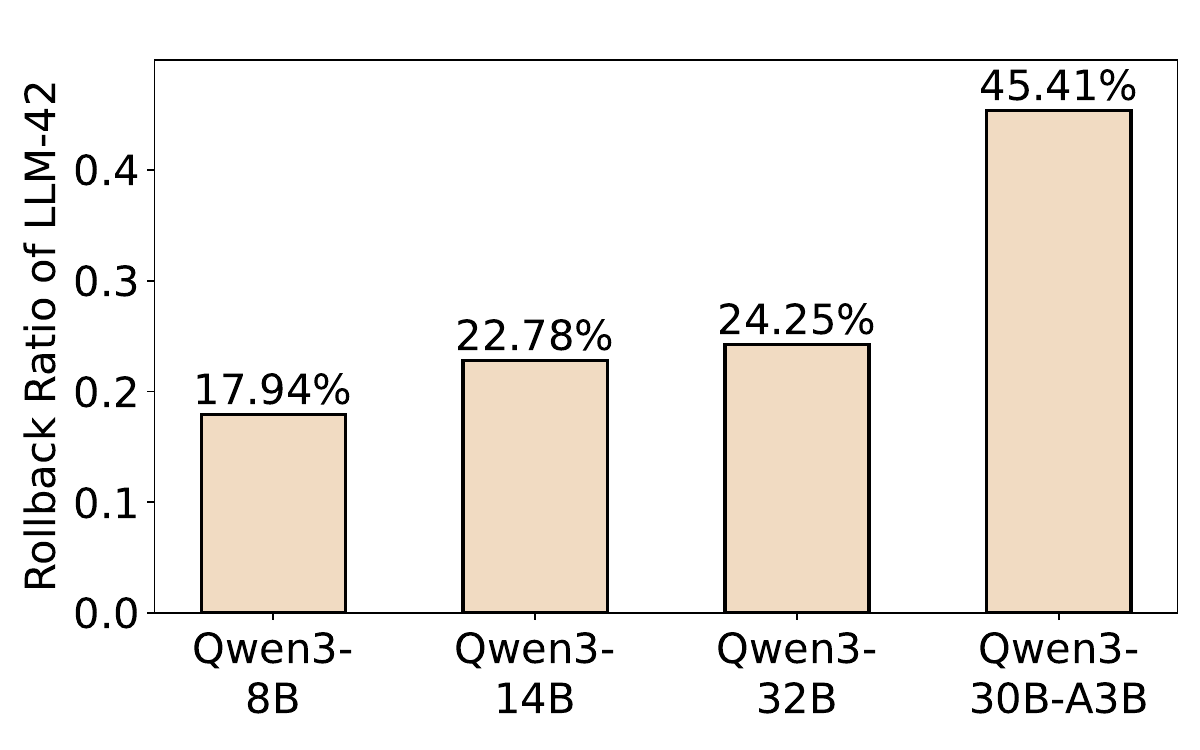}
        \caption{Rollback ratio across model sizes}
        \label{fig:llm42b}
    \end{subfigure}
    \caption{Applicability and rollback cost of LLM-42. The verification
    window is 64 tokens in both experiments.}
    \label{fig:llm42}
\end{figure}

\figurename~\ref{fig:llm42a} compares throughput across concurrency levels and
deterministic-request fractions. ``LLM-42 @ $k$\%'' denotes a workload in
which $k$\% of requests require determinism; we set its verification window
to 64 tokens. At a low deterministic fraction, LLM-42 can approach or
occasionally exceed \sys\ because most requests avoid verification. As the
fraction grows, verification becomes frequent and its throughput falls. When
all requests require determinism, as in model evaluation and RL rollout,
LLM-42 performs below both \sys\ and the BI baseline in our experiment.

The verification rate explains this trend. At concurrency 256, each decode
iteration produces 256 candidate tokens. With a 64-token window, an average
of $256/64=4$ requests become eligible for verification per iteration. Each
verification replays a fixed-shape model step, which is substantially more
expensive than generating tokens on a fast dynamic path.
Frequent verification therefore dominates execution when deterministic
traffic is dense. \sys\ is better suited to this all-deterministic,
high-concurrency regime.

\subsection{Variable-Concurrency Online Serving}

LLM-42 is also better suited to online services whose request rate is unknown
or changes over time. \sys\ must choose a maximum concurrency before launch
because that value determines the padded decode shape. Setting it too high
adds many padding tokens when traffic is light and reduces decode throughput.
Setting it too low leaves requests waiting outside the running batch even
when the system could exploit more parallelism. No single fixed setting is
efficient across a wide range of dynamically changing arrival rates. By
retaining dynamic batching on its fast path, LLM-42 adapts to variable
concurrency without this configuration constraint.

\subsection{Rollback Cost of Verified Speculation}
Our experiments also reveal a limitation of LLM-42.
\figurename~\ref{fig:llm42b} reports its rollback ratio on
ShareGPT~\cite{sharegpt} with a 64-token verification window. We define the
rollback ratio as the number of rejected candidate tokens divided by the
total number of tokens generated on the nondeterministic fast path. A larger
ratio represents more wasted computation and a higher verification penalty.

The rollback ratio rises from 17.94\% on Qwen3-8B to 24.25\% on Qwen3-32B,
suggesting that deeper model execution provides more opportunities for small
numerical differences to affect a token decision. Qwen3-30B-A3B, a
Mixture-of-Experts (MoE) model, reaches 45.41\%, nearly twice the ratio of the
similarly sized dense Qwen3-32B. One plausible explanation is that a small router perturbation can change the
selected experts, after which the two execution paths perform different
computations rather than merely accumulating different low-order bits. High
rollback rates may therefore limit verified speculation on large MoE models.

\section{Future Work}

The ablation study identifies isolated prefill as \sys's largest source of
overhead. We consider two directions for reducing this cost.

\subsection{Batch-Invariant Kernels for Prefill}

\sys\ isolates prefill because mixing prefill and decode would destroy the
fixed decode shape. Prefill requests could still be batched with other
prefill requests, which would substantially reduce the cost of short prompts.
Such batching changes the model input shape, so the prefill stage would need
batch-invariant kernels. A hybrid design could use batch-invariant kernels for
batched prefill and faster position-invariant kernels for fixed-shape decode.
Its performance relative to isolated prefill requires experimental study, and
maintaining different kernel paths for the two stages introduces an
engineering challenge.

\subsection{Prefill-Decode Disaggregation}

\sys's separation of prefill and decode maps naturally to existing
prefill-decode disaggregation systems~\cite{patel2024,qin2025mooncake,
zhong2024}. Prefill and decode instances can use different parallelization
and kernel configurations. Decode instances can apply \sys's fixed-shape
execution without modification, while the deployment can provision more
prefill instances to prevent isolated prefill from limiting end-to-end
throughput. Disaggregation also simplifies the hybrid-kernel design above:
prefill instances can use batch-invariant kernels, while decode instances
retain optimized position-invariant kernels.

\section{Conclusion}

This paper shows that full batch invariance is
unnecessary when optimized kernels are position-invariant and the serving
system controls their execution shapes. \sys\ applies this insight through
isolated prefill, fixed-shape CUDA-Graph decode, selective constraints on
Split-KV and collective reductions, and request-bound sampling state. We
formalize this guarantee through an end-to-end determinism contract and an
inductive argument showing that request state and sampling decisions remain
unchanged across co-runners, batch positions, and arrival orders. Across three
large models and diverse workloads, \sys\ preserves deterministic token
generation while improving throughput by 15--324\% over batch-invariant
inference. \sys\ is most effective for high-concurrency workloads in which
all requests require deterministic outputs. These results establish
execution-shape control as a practical alternative to making the entire LLM
stack batch-invariant.

\bibliographystyle{IEEEtranS}
\bibliography{refs}

\begin{thebibliography}{10}
\providecommand{\url}[1]{#1}
\csname url@samestyle\endcsname
\providecommand{\newblock}{\relax}
\providecommand{\bibinfo}[2]{#2}
\providecommand{\BIBentrySTDinterwordspacing}{\spaceskip=0pt\relax}
\providecommand{\BIBentryALTinterwordstretchfactor}{4}
\providecommand{\BIBentryALTinterwordspacing}{\spaceskip=\fontdimen2\font plus
\BIBentryALTinterwordstretchfactor\fontdimen3\font minus \fontdimen4\font\relax}
\providecommand{\BIBforeignlanguage}[2]{{%
\expandafter\ifx\csname l@#1\endcsname\relax
\typeout{** WARNING: IEEEtranS.bst: No hyphenation pattern has been}%
\typeout{** loaded for the language `#1'. Using the pattern for}%
\typeout{** the default language instead.}%
\else
\language=\csname l@#1\endcsname
\fi
#2}}
\providecommand{\BIBdecl}{\relax}
\BIBdecl

\bibitem{sharegpt}
``Sharegpt\_vicuna\_unfiltered,'' \url{https://huggingface.co/datasets/anon8231489123/ShareGPT_Vicuna_unfiltered}, 2023.

\bibitem{agrawal2024sarathi}
\BIBentryALTinterwordspacing
A.~Agrawal, N.~Kedia, A.~Panwar, J.~Mohan, N.~Kwatra, B.~Gulavani, A.~Tumanov, and R.~Ramjee, ``Taming {Throughput-Latency} tradeoff in {LLM} inference with {Sarathi-Serve},'' in \emph{18th USENIX Symposium on Operating Systems Design and Implementation (OSDI 24)}.\hskip 1em plus 0.5em minus 0.4em\relax Santa Clara, CA: USENIX Association, Jul. 2024, pp. 117--134. [Online]. Available: \url{https://www.usenix.org/conference/osdi24/presentation/agrawal}
\BIBentrySTDinterwordspacing

\bibitem{anadkat2023seed}
S.~Anadkat, \emph{How to make your completions outputs consistent with the new seed parameter}, \url{https://developers.openai.com/cookbook/examples/reproducible_outputs_with_the_seed_parameter}, 2023.

\bibitem{atil2025non}
\BIBentryALTinterwordspacing
B.~At{\i}l, S.~Aykent, A.~Chittams, L.~Fu, R.~J. Passonneau, E.~Radcliffe, G.~R. Rajagopal, A.~Sloan, T.~Tudrej, F.~Ture, Z.~Wu, L.~Xu, and B.~Baldwin, ``Non-determinism of ``deterministic'' {LLM} system settings in hosted environments,'' in \emph{Proceedings of the 5th Workshop on Evaluation and Comparison of NLP Systems}, M.~Akter, T.~Chowdhury, S.~Eger, C.~Leiter, J.~Opitz, and E.~{\c{C}}ano, Eds.\hskip 1em plus 0.5em minus 0.4em\relax Mumbai, India: Association for Computational Linguistics, Dec. 2025, pp. 135--148. [Online]. Available: \url{https://aclanthology.org/2025.eval4nlp-1.12/}
\BIBentrySTDinterwordspacing

\bibitem{chann2023non}
S.~Chann, ``Non-determinism in gpt-4 is caused by sparse moe,'' \url{https://152334h.github.io/blog/non-determinism-in-gpt-4/}, 2023.

\bibitem{chu2026margingate}
\BIBentryALTinterwordspacing
K.~Chu, Y.~Zhou, and W.~Zhang, ``Margingate: Sparse margin-triggered verification for batch-invariant llm inference,'' 2026. [Online]. Available: \url{https://arxiv.org/abs/2605.30218}
\BIBentrySTDinterwordspacing

\bibitem{daita2025blazedit}
V.~Daita, X.~Lian, L.~Zhang, and J.~Liu, ``Blazing-fast code editing via multi-layer speculation,'' \url{https://github.com/ise-uiuc/blazedit}, 2025.

\bibitem{dao2024flashattention2}
\BIBentryALTinterwordspacing
T.~Dao, ``Flashattention-2: Faster attention with better parallelism and work partitioning,'' in \emph{International Conference on Learning Representations}, B.~Kim, Y.~Yue, S.~Chaudhuri, K.~Fragkiadaki, M.~Khan, and Y.~Sun, Eds., vol. 2024, 2024, pp. 35\,549--35\,562. [Online]. Available: \url{https://proceedings.iclr.cc/paper_files/paper/2024/file/98ed250b203d1ac6b24bbcf263e3d4a7-Paper-Conference.pdf}
\BIBentrySTDinterwordspacing

\bibitem{dao2022flashattention}
\BIBentryALTinterwordspacing
T.~Dao, D.~Fu, S.~Ermon, A.~Rudra, and C.~R\'{e}, ``Flashattention: Fast and memory-efficient exact attention with io-awareness,'' in \emph{Advances in Neural Information Processing Systems}, S.~Koyejo, S.~Mohamed, A.~Agarwal, D.~Belgrave, K.~Cho, and A.~Oh, Eds., vol.~35.\hskip 1em plus 0.5em minus 0.4em\relax Curran Associates, Inc., 2022, pp. 16\,344--16\,359. [Online]. Available: \url{https://proceedings.neurips.cc/paper_files/paper/2022/file/67d57c32e20fd0a7a302cb81d36e40d5-Paper-Conference.pdf}
\BIBentrySTDinterwordspacing

\bibitem{dao2023flashdecoding}
T.~Dao, D.~Haziza, F.~Massa, and G.~Sizov, ``Flash-decoding for long-context inference,'' \url{https://crfm.stanford.edu/2023/10/12/flashdecoding.html}, 2023.

\bibitem{deepseekai2024deepseekv3technicalreport}
\BIBentryALTinterwordspacing
DeepSeek-AI, ``Deepseek-v3 technical report,'' 2024. [Online]. Available: \url{https://arxiv.org/abs/2412.19437}
\BIBentrySTDinterwordspacing

\bibitem{deepseekai2026deepseekv4}
\BIBentryALTinterwordspacing
DeepSeek-AI, A.~Xu, B.~Lin, B.~Xue, B.~Wang, B.~Xu, B.~Wu, B.~Zhang, C.~Lin, C.~Dong, C.~Ling, C.~Lu, C.~Zhao, C.~Deng, C.~Hou, C.~Xu, C.~Shao, C.~Ruan, C.~Sun, D.~Dai, D.~Guo, D.~Yang, D.~Chen, D.~Li, D.~Ji, E.~Li, F.~Wei, F.~Lin, F.~Yuan, F.~Xia, F.~Dai, G.~Hao, G.~Chen, G.~Cao, G.~Meng, G.~Li, H.~Yu, H.~Zhang, H.~Xu, H.~Li, H.~Liang, H.~Zhang, H.~Luo, H.~Wei, H.~Yuan, H.~Zhang, H.~Luo, H.~Chen, H.~Ji, H.~Zhang, H.~Ding, H.~Tang, H.~Cao, H.~Gao, H.~Qu, H.~Zeng, J.~Yang, J.~Zhu, J.~Luo, J.~Song, J.~Yu, J.~Huang, J.~Cai, J.~Liang, J.~Zhou, J.~Ye, J.~Li, J.~Xu, J.~Hu, J.~Yang, J.~Chen, J.~Yan, J.~Chen, J.~Zhou, J.~Xiang, J.~Yuan, J.~Cheng, J.~Zhou, J.~Zhu, J.~Yu, J.~Sun, J.~Ran, J.~Jiang, J.~Qiu, J.~Li, J.~Zheng, J.~Song, K.~Dong, K.~Gao, K.~Guan, K.~Zhou, K.~Huang, K.~Yu, L.~Wang, L.~Zhang, L.~Wang, L.~Xia, L.~Zhang, L.~Zhao, L.~Guo, L.~Luo, L.~Ma, L.~Zhu, L.~Wang, L.~Cai, L.~Zhang, L.~Chen, M.~Di, M.~Xu, M.~Mei, M.~Wang, M.~Zhang, M.~Zhang, M.~Tang, M.~Li, M.~Zhou, M.~Han, N.~Wang, P.~Huang, P.~Wang,
  P.~Cong, P.~Wang, P.~Zhang, Q.~Wang, Q.~Zhu, Q.~Li, Q.~Chen, Q.~Du, Q.~Jiang, R.~Tian, R.~Xu, R.~Lu, R.~Xu, R.~Ge, R.~Zhang, R.~Pan, R.~Wang, R.~Chen, R.~Yin, R.~Xu, R.~Shen, R.~Zhang, R.~Chen, S.~Liu, S.~Lu, S.~Sun, S.~Zhou, S.~Chen, S.~Cai, S.~Nie, S.~Wu, S.~Chen, S.~Hu, S.~Liu, S.~Hu, S.~Ma, S.~Wang, S.~Yu, S.~Zhou, S.~Pan, S.~Yu, S.~Zhou, T.~Ni, T.~Yun, T.~Jin, T.~Pei, T.~Ye, T.~Lin, T.~Ji, T.~Cui, T.~Yue, T.~Yu, T.~Wang, W.~Zhang, W.~Xiao, W.~Zeng, W.~An, W.~Zhao, W.~Liu, W.~Liang, W.~Pang, W.~Luo, W.~Yao, W.~Gao, W.~Yang, W.~Huang, W.~Hou, W.~Zhang, W.~Ma, X.~Gao, X.~He, X.~Wang, X.~Wang, X.~Bi, X.~Liu, X.~Wang, X.~Chen, X.~Zhang, X.~Nie, X.~Sun, X.~Wang, X.~Cheng, X.~Liu, X.~Xie, X.~Liu, X.~Liu, X.~Yu, X.~Li, X.~Yang, X.~Zhang, X.~Chen, X.~Wang, X.~Su, X.~Chen, X.~Lin, X.~Fu, Y.~Yan, Y.~Wang, Y.~Ma, Y.~Luo, Y.~Zhang, Y.~Xu, Y.~Ma, Y.~Huang, Y.~Li, Y.~Li, Y.~Xu, Y.~Zhao, Y.~Sun, Y.~Wang, Y.~Qian, Y.~Shao, Y.~Yu, Y.~Zhang, Y.~Ding, Y.~Shi, Y.~Wu, Y.~Xiong, Y.~Ma, Y.~He, Y.~Tang, Y.~Zhou, Y.~Luo,
  Y.~Zhong, Y.~Piao, Y.~Wang, Y.~Zhang, Y.~Chen, Y.~Tan, Y.~Wei, Y.~Ma, Y.~Liu, Y.~Yang, Y.~Guo, Y.~Wu, Y.~Wu, Y.~Li, Y.~Cheng, Y.~Ou, Y.~Xu, Y.~Li, Y.~Wang, Y.~Yang, Y.~Xu, Y.~Wu, Y.~Meng, Y.~Zou, Y.~Zha, Y.~Xiong, Y.~Chen, Y.~Lin, Y.~Cao, Y.~Wang, Y.~Zhang, Y.~Yan, Y.~Lin, Y.~Gu, Y.~Luo, Y.~You, Y.~Liu, Y.~Zhou, Y.~Zhou, Y.~Huang, Z.~Wu, Z.~Wang, Z.~Zhao, Z.~Ren, Z.~Zhang, Z.~Sha, Z.~Fu, Z.~Ju, Z.~Xu, Z.~Xie, Z.~Zhang, Z.~Gao, Z.~Hao, Z.~Gou, Z.~Ma, Z.~Yan, Z.~Shao, Z.~Huang, Z.~Chen, Z.~Wu, Z.~Ren, Z.~Wu, Z.~Li, Z.~Zhang, Z.~Xu, Z.~Wang, Z.~Qu, Z.~Gu, Z.~Zhu, Z.~Li, Z.~Zhang, Z.~Xie, Z.~Gao, Z.~Wan, Z.~Pan, and Z.~Yao, ``Deepseek-v4: Towards highly efficient million-token context intelligence,'' 2026. [Online]. Available: \url{https://arxiv.org/abs/2606.19348}
\BIBentrySTDinterwordspacing

\bibitem{goldberg1991floating}
\BIBentryALTinterwordspacing
D.~Goldberg, ``What every computer scientist should know about floating-point arithmetic,'' \emph{ACM Comput. Surv.}, vol.~23, no.~1, p. 5–48, Mar. 1991. [Online]. Available: \url{https://doi.org/10.1145/103162.103163}
\BIBentrySTDinterwordspacing

\bibitem{gond2026llm42}
\BIBentryALTinterwordspacing
R.~Gond, A.~K. Kamath, R.~Ramjee, and A.~Panwar, ``Llm-42: Enabling determinism in llm inference with verified speculation,'' 2026. [Online]. Available: \url{https://arxiv.org/abs/2601.17768}
\BIBentrySTDinterwordspacing

\bibitem{he2025nondeterminism}
H.~He and T.~M. Lab, ``Defeating nondeterminism in llm inference,'' \emph{Thinking Machines Lab: Connectionism}, 2025, https://thinkingmachines.ai/blog/defeating-nondeterminism-in-llm-inference/.

\bibitem{kwon2023efficient}
\BIBentryALTinterwordspacing
W.~Kwon, Z.~Li, S.~Zhuang, Y.~Sheng, L.~Zheng, C.~H. Yu, J.~Gonzalez, H.~Zhang, and I.~Stoica, ``Efficient memory management for large language model serving with pagedattention,'' in \emph{Proceedings of the 29th Symposium on Operating Systems Principles}, ser. SOSP '23.\hskip 1em plus 0.5em minus 0.4em\relax New York, NY, USA: Association for Computing Machinery, 2023, p. 611–626. [Online]. Available: \url{https://doi.org/10.1145/3600006.3613165}
\BIBentrySTDinterwordspacing

\bibitem{nvidia2026cudagraphs}
\BIBentryALTinterwordspacing
{NVIDIA Corporation}, \emph{CUDA Programming Guide: CUDA Graphs}, 2026. [Online]. Available: \url{https://docs.nvidia.com/cuda/cuda-programming-guide/04-special-topics/cuda-graphs.html}
\BIBentrySTDinterwordspacing

\bibitem{patel2024}
P.~Patel, E.~Choukse, C.~Zhang, A.~Shah, I.~Goiri, S.~Maleki, and R.~Bianchini, ``Splitwise: Efficient generative llm inference using phase splitting,'' in \emph{2024 ACM/IEEE 51st Annual International Symposium on Computer Architecture (ISCA)}, 2024, pp. 118--132.

\bibitem{qin2025mooncake}
\BIBentryALTinterwordspacing
R.~Qin, Z.~Li, W.~He, J.~Cui, F.~Ren, M.~Zhang, Y.~Wu, W.~Zheng, and X.~Xu, ``Mooncake: Trading more storage for less computation {\textemdash} a {KVCache-centric} architecture for serving {LLM} chatbot,'' in \emph{23rd USENIX Conference on File and Storage Technologies (FAST 25)}.\hskip 1em plus 0.5em minus 0.4em\relax Santa Clara, CA: USENIX Association, Feb. 2025, pp. 155--170. [Online]. Available: \url{https://www.usenix.org/conference/fast25/presentation/qin}
\BIBentrySTDinterwordspacing

\bibitem{rainbird2025}
{Rainbird AI}, ``Deterministic graph-based inference: The key to safe ai in financial services,'' \url{https://rainbird.ai/deterministic-graph-based-inference-the-key-to-safe-ai-in-financial-services/}, 2025.

\bibitem{sglang2025det}
{SGLang Team}, ``Towards deterministic inference in sglang and reproducible rl training,'' \url{https://www.lmsys.org/blog/2025-09-22-sglang-deterministic/}, 2025.

\bibitem{tencent2026hy3}
{Tencent Hunyuan Team}, ``{Tencent Hy3},'' \url{https://huggingface.co/tencent/Hy3}, 2026.

\bibitem{vllm2026det}
vLLM Team, \emph{Batch Invariance}, \url{https://docs.vllm.ai/en/latest/features/batch_invariance/}, 2026.

\bibitem{vllm2026cudagraph}
vLLM Team, \emph{CUDA Graphs}, \url{https://docs.vllm.ai/en/latest/design/cuda_graphs/}, 2026.

\bibitem{vllm2026test}
vLLM Team, ``test\_batch\_invariance,'' \url{https://github.com/vllm-project/vllm/blob/main/tests/v1/determinism/test_batch_invariance.py}, 2026.

\bibitem{wang2025burstgpt}
\BIBentryALTinterwordspacing
Y.~Wang, Y.~Chen, Z.~Li, X.~Kang, Y.~Fang, Y.~Zhou, Y.~Zheng, Z.~Tang, X.~He, R.~Guo, X.~Wang, Q.~Wang, A.~C. Zhou, and X.~Chu, ``Burstgpt: A real-world workload dataset to optimize llm serving systems,'' in \emph{Proceedings of the 31st ACM SIGKDD Conference on Knowledge Discovery and Data Mining V.2}, ser. KDD '25.\hskip 1em plus 0.5em minus 0.4em\relax New York, NY, USA: Association for Computing Machinery, 2025, p. 5831–5841. [Online]. Available: \url{https://doi.org/10.1145/3711896.3737413}
\BIBentrySTDinterwordspacing

\bibitem{qwen3technicalreport}
\BIBentryALTinterwordspacing
A.~Yang, A.~Li, B.~Yang, B.~Zhang, B.~Hui, B.~Zheng, B.~Yu, C.~Gao, C.~Huang, C.~Lv, C.~Zheng, D.~Liu, F.~Zhou, F.~Huang, F.~Hu, H.~Ge, H.~Wei, H.~Lin, J.~Tang, J.~Yang, J.~Tu, J.~Zhang, J.~Yang, J.~Yang, J.~Zhou, J.~Zhou, J.~Lin, K.~Dang, K.~Bao, K.~Yang, L.~Yu, L.~Deng, M.~Li, M.~Xue, M.~Li, P.~Zhang, P.~Wang, Q.~Zhu, R.~Men, R.~Gao, S.~Liu, S.~Luo, T.~Li, T.~Tang, W.~Yin, X.~Ren, X.~Wang, X.~Zhang, X.~Ren, Y.~Fan, Y.~Su, Y.~Zhang, Y.~Zhang, Y.~Wan, Y.~Liu, Z.~Wang, Z.~Cui, Z.~Zhang, Z.~Zhou, and Z.~Qiu, ``Qwen3 technical report,'' 2025. [Online]. Available: \url{https://arxiv.org/abs/2505.09388}
\BIBentrySTDinterwordspacing

\bibitem{feng2025offpolicy}
\BIBentryALTinterwordspacing
F.~Yao, L.~Liu, D.~Zhang, C.~Dong, J.~Shang, and J.~Gao, ``Your efficient rl framework secretly brings you off-policy rl training,'' Aug. 2025. [Online]. Available: \url{https://fengyao.notion.site/off-policy-rl}
\BIBentrySTDinterwordspacing

\bibitem{yu2022orca}
\BIBentryALTinterwordspacing
G.-I. Yu, J.~S. Jeong, G.-W. Kim, S.~Kim, and B.-G. Chun, ``Orca: A distributed serving system for {Transformer-Based} generative models,'' in \emph{16th USENIX Symposium on Operating Systems Design and Implementation (OSDI 22)}.\hskip 1em plus 0.5em minus 0.4em\relax Carlsbad, CA: USENIX Association, Jul. 2022, pp. 521--538. [Online]. Available: \url{https://www.usenix.org/conference/osdi22/presentation/yu}
\BIBentrySTDinterwordspacing

\bibitem{yuan2025fp32}
\BIBentryALTinterwordspacing
J.~Yuan, H.~Li, X.~Ding, W.~Xie, Y.-J. Li, W.~Zhao, K.~Wan, J.~Shi, X.~Hu, and Z.~Liu, ``Understanding and mitigating numerical sources of nondeterminism in llm inference,'' in \emph{Advances in Neural Information Processing Systems}, D.~Belgrave, C.~Zhang, H.~Lin, R.~Pascanu, P.~Koniusz, M.~Ghassemi, and N.~Chen, Eds., vol.~38.\hskip 1em plus 0.5em minus 0.4em\relax Curran Associates, Inc., 2025, pp. 169\,819--169\,851. [Online]. Available: \url{https://proceedings.neurips.cc/paper_files/paper/2025/file/f80094a824ba5912d4a2de169c404a40-Paper-Conference.pdf}
\BIBentrySTDinterwordspacing

\bibitem{zhang2026deterministic}
\BIBentryALTinterwordspacing
Z.~Zhang, X.~Ding, J.~Yuan, R.~Liu, H.~Mao, J.~Xing, and Z.~Liu, ``Deterministic inference across tensor parallel sizes that eliminates training-inference mismatch,'' in \emph{Forty-third International Conference on Machine Learning}, 2026. [Online]. Available: \url{https://openreview.net/forum?id=5eZmlUyFpl}
\BIBentrySTDinterwordspacing

\bibitem{zheng2024sglang}
\BIBentryALTinterwordspacing
L.~Zheng, L.~Yin, Z.~Xie, C.~Sun, J.~Huang, C.~H. Yu, S.~Cao, C.~Kozyrakis, I.~Stoica, J.~E. Gonzalez, C.~Barrett, and Y.~Sheng, ``Sglang: Efficient execution of structured language model programs,'' in \emph{Advances in Neural Information Processing Systems}, A.~Globerson, L.~Mackey, D.~Belgrave, A.~Fan, U.~Paquet, J.~Tomczak, and C.~Zhang, Eds., vol.~37.\hskip 1em plus 0.5em minus 0.4em\relax Curran Associates, Inc., 2024, pp. 62\,557--62\,583. [Online]. Available: \url{https://proceedings.neurips.cc/paper_files/paper/2024/file/724be4472168f31ba1c9ac630f15dec8-Paper-Conference.pdf}
\BIBentrySTDinterwordspacing

\bibitem{zhong2026diagnosing}
\BIBentryALTinterwordspacing
T.~Zhong, N.~Ling, Y.~Pi, Z.~Wei, T.~Yu, G.~Fox, P.~Wu, and X.~Yu, ``Diagnosing training inference mismatch in llm reinforcement learning,'' 2026. [Online]. Available: \url{https://arxiv.org/abs/2605.14220}
\BIBentrySTDinterwordspacing

\bibitem{zhong2024}
\BIBentryALTinterwordspacing
Y.~Zhong, S.~Liu, J.~Chen, J.~Hu, Y.~Zhu, X.~Liu, X.~Jin, and H.~Zhang, ``{DistServe}: Disaggregating prefill and decoding for goodput-optimized large language model serving,'' in \emph{18th USENIX Symposium on Operating Systems Design and Implementation (OSDI 24)}.\hskip 1em plus 0.5em minus 0.4em\relax Santa Clara, CA: USENIX Association, Jul. 2024, pp. 193--210. [Online]. Available: \url{https://www.usenix.org/conference/osdi24/presentation/zhong-yinmin}
\BIBentrySTDinterwordspacing

\end{thebibliography}

\end{document}